\documentclass[aps,twocolumn,footinbib,showpacs,superscriptaddress,floatfix,bibliography,prb]{revtex4-1}

\usepackage{times}
\usepackage{float}
\usepackage[dvipsnames]{xcolor}
\usepackage{amsmath}
\usepackage{amsthm}
\usepackage{amssymb}
\usepackage{amsbsy}
\usepackage{enumitem}
\usepackage{wasysym}
\usepackage[english]{babel}
\usepackage[T1]{fontenc}
\usepackage[utf8]{inputenc} 
\usepackage{graphicx}
\usepackage{ulem}
\usepackage{pstricks}
\usepackage{rotating}			       
\usepackage{tabularx,hhline}	
\usepackage[caption=false]{subfig}		

\usepackage{setspace}
\usepackage{array,tabularx}
\usepackage{appendix}
\usepackage{booktabs}
\usepackage{graphicx}
\usepackage{epstopdf}
\usepackage{color}
\usepackage[colorlinks,bookmarks=false,citecolor=blue,linkcolor=red,urlcolor=blue]{hyperref}
\usepackage{tikz}
\usetikzlibrary{calc}
\usepackage{multirow}
\usepackage{bm}

\graphicspath{{Figures/}}

\usetikzlibrary{shapes}

\DeclareRobustCommand{\singAB}{
\begin{tikzpicture}[baseline={([yshift=-1.5]current bounding box.center)}]]
    \filldraw[color=black, fill=black, thick](-8,0) circle (0.04);
    \filldraw[color=black, fill=black, thick](-8.,0.2) circle (0.04);
    \filldraw[color=black,fill=white,thick](-8,0.04)  rectangle (-8.,0.24);
     \end{tikzpicture}
     }
\newcommand{\figref}[1]{Fig.~\ref{fig:#1}}

\newcommand{\Secref}[1]{Sec.~\ref{sec:#1}}

\newcolumntype{P}[1]{>{\centering\arraybackslash}p{#1}}

\DeclareMathOperator{\sgn}{sgn}

\begin{document}

\title{Fragility of $\mathcal{Z}_2$ topological invariant characterizing triplet excitations in a bilayer kagome magnet}

\author{Andreas Thomasen}
\affiliation{Theory of Quantum Matter Unit, Okinawa Institute of Science and Technology Graduate University, Onna-son, Okinawa 904-0395, Japan}

\author{Karlo Penc}
\affiliation{Institute for Solid State Physics and Optics, Wigner Research Centre for Physics, H-1525 Budapest, P.O.B. 49, Hungary}

\author{Nic Shannon}
\affiliation{Theory of Quantum Matter Unit, Okinawa Institute of Science and Technology Graduate University, Onna-son, Okinawa 904-0395, Japan}

\author{Judit Romh\'anyi}
\affiliation{Theory of Quantum Matter Unit, Okinawa Institute of Science and Technology Graduate University, Onna-son, Okinawa 904-0395, Japan}
\affiliation{Department of Physics and Astronomy, University of California, Irvine, California 92697, USA}

\date{\today}

\begin{abstract}

The discovery by Kane and Mele of a model of spinful electrons characterized by a $\mathcal{Z}_2$ topological invariant had a lasting effect on the study of electronic band structures.
Given this, it is natural to ask whether similar topology can be found in the band-like excitations of magnetic insulators, and recently models supporting $\mathcal{Z}_2$ topological invariants have been proposed for both magnon [Kondo et al. Phys. Rev. B 99, 041110(R) (2019)] and triplet [D. G. Joshi and A. P. Schnyder, Phys. Rev. B 100, 020407 (2019)] excitations.
In both cases, magnetic excitations form time-reversal (TR) partners, which mimic the Kramers pairs of electrons in the Kane-Mele model but do not enjoy the same type of symmetry protection. 
In this paper, we revisit this problem in the context of the triplet excitations of a spin model on the bilayer kagome lattice. 
Here the triplet excitations provide a faithful analog of the Kane–Mele model as long as the Hamiltonian preserves the TR$\times$U(1) symmetry.
We find that exchange anisotropies, allowed by the point group and typical in realistic models, break the required TR$\times$U(1) symmetry and instantly destroy the $\mathcal{Z}_2$ band topology. 
We further consider the effects of TR breaking by an applied magnetic field. 
In this case, the lifting of spin–degeneracy leads to a triplet Chern insulator, which is stable against the breaking of TR$\times$U(1) symmetry. 
Kagome bands realize both a quadratic and a linear band touching, and we provide a thorough characterization of the Berry curvature associated with both cases. 
We also calculate the triplet–mediated spin Nernst and thermal Hall signals which could be measured in experiments.
These results suggest that the $\mathcal{Z}_2$ topology of band--like excitations in magnets may be intrinsically fragile compared to their electronic counterparts.

\end{abstract}

\maketitle

\section{Introduction}\label{sec:topologicalTriplons}

The discovery, in 1980, of an integer quantization in the Hall response 
of a two--dimensional electron gas in high magnetic field \cite{Klitzing1980}, 
marked a new beginning for condensed matter physics.
It was quickly realised that this quantization implied a new form of 
universality \cite{Laughlin1981}, enjoying protection against both interactions and disorder, 
with the quantized Hall response reflecting the integer 
values of the Chern indices characterizing the topology of the underlying 
electron bands \cite{Thouless1982,Avron1983,Kohmoto1985}.
In a celebrated paper, Haldane noted that these conditions could also be met  
in a simple model of spinless electrons on a honeycomb lattice, with 
time--reversal symmetry broken by complex hoping integrals, 
but no magnetic field \cite{Haldane1988}.
And the generalization of Haldane's model to spinful electrons, 
by Kane and Mele \cite{Kane2005,Kane2005b}, set the stage for the 
burgeoning field of topological insulators (TI's) and superconductors \cite{Hasan2010,Qi2011}, 
with current estimates suggesting that as many as 27\% of materials may 
have a topological band structure \cite{Vergniory2019}.   
Moreover, since the exotic properties of TI's follow from the single--particle 
properties of a band, analagous effects can also be found in a wide 
range of other systems, 
including photonic metastructures \cite{Onoda2004,Ozawa2019}, electronic circuits \cite{Lee2018}, 
and both accoustic \cite{Peri2019} and mechanical lattices \cite{Fruchart2020}.

Another natural places to look for nontrivial topology is in the band-like integer spin 
excitations of insulating magnets.
These may take the form of magnon (spin--wave) excitations of 
ordered phases, or triplet excitations of quantum paramagnets.
Such excitations are Bosonic, and acquire Berry phases as a consequence 
of spin--orbit coupling, usually in the form of Dzyaloshniski--Moriya 
interactions \cite{Onose2010}.
As a result, both magnon 
\cite{Shindou2013,Mook2014,Mook2014b,Owerre2016,Chernyshev2016,Zyuzin2016,Kim2016,Nakata2017,McClarty2018,Kondo2019,McClarty2019,Kondo2020,McClarty-arXiv}, 
and triplon \cite{Romhanyi2015,McClarty2017,Joshi2019}, 
bands can exhibit non-trivial Chern indices, in direct analogue with TI's~\cite{Ganesh2020}.
These systems exhibit exactly the same topologically--protected edge modes as  
their electronic counterparts \cite{Shindou2013,Mook2014b,Romhanyi2015,Nakata2017}, and can be indexed 
in the same way, even in the presence of disorder \cite{Akagi2017,Yoshioka2018}.
They also support thermal Hall \cite{Mook2014,Romhanyi2015,Owerre2016,Nakata2017} and spin--Nernst \cite{Zyuzin2016,Kim2016,Cheng2016,Owerre2016,Nakata2017,Kondo2019} 
effects, in correspondence to the Hall~\cite{Haldane1988} 
and spin--Hall effects seen in electronic systems \cite{Kane2005,Kane2005b}.

However, since these topological bands are a feature of excitations, rather than 
of a ground state, the quantized Hall effect found in Chernful band of electrons \cite{Thouless1982,Avron1983,Kohmoto1985}, 
is superseded by a non--integer, temperature--dependent 
response, coming from thermally--excited Bosons \cite{Katsura2010,Matsumoto2011}.
Moreover, the fact that interactions between Bosons can be relevant \cite{Chernyshev2015,Chernyshev2016}
also creates a new opportunities to study non--Hermitian aspects of their 
dynamics \cite{McClarty2019,Kondo2020}.
For a recent review of this, and other related issues see \cite{McClarty-arXiv}.

Given the seminal role of the models of Haldane \cite{Haldane1988}, 
and Kane and Mele \cite{Kane2005,Kane2005b} 
in the understanding of electronic TI's, 
it is natural to look for corresponding systems in magnets.
The route to a Haldane model for magnons turns out to be both simple and elegant:
the Heisenberg ferromagnet (FM) on a honeycomb lattice realises magnons with a 
graphene--like dispersion \cite{Franson2016}, and the symmetry of this lattice 
permits DM interactions on second--neighbour bonds.
These supply the complex hopping integral invoked by Haldane, 
opening a gap in the magnon dispersion, and endowing the bands with 
Chern numbers \cite{Kim2016,Kim2017}.  
By extension, an exact analogue of the Kane--Mele model can be realised 
in a bilayer honeycomb magnet, with interlayer interactions 
chosen such that it forms two copies of a Haldane model, with magnon bands 
related by time--reversal symmetry \cite{Kondo2019}, 
an approach which can be extended to the 
Fu--Kane--Mele model in three dimensions \cite{Kondo2019-PRB100}.
It is also possible to achieve triplon bands which mirror the Kane--Mele 
model, in a quantum paramagnet on a bilayer honeycomb lattice \cite{Joshi2019}.


While the route to topological bands in magnets is now well established, 
a number of important questions remain.
In particular, most work to date has taken a ``top--down'' approach, emphasizing 
how topological effects  found in electronic systems can be recreated within the 
band--like excitations of magnets.
Less attention has been paid to features which may be unique to magnets, 
or to building models of topological phases in magnets from the ``ground up'', 
starting from the most general spin interactions allowed by symmetry of a 
given lattice.
One risk inherent in the ``top down'' approach
is that the symmetries which protect topological phases formed by electrons, 
need not protect those formed by spins.
For example, while the $S^z = \pm1$ triplet excitations of a quantum 
paramagnet form a doublet under time--reversal symmetry, they do not 
satisfy Kramer's theorem.
This means that the $S^z = \pm1$ doublets are much more ``fragile'' than the 
electronic doublets considered by Kane and Mele, and the consequences of 
any symmetry--allowed terms in the Hamiltonian which mix triplets with 
$S^z = \pm1$, therefore need to be considered explicitly.
It is also of interest to ask what such a topological quantum paramagnet would 
look like in experiment, and how this physics might generalize to structures more 
complicated than the honeycomb lattice.


In this Article, we address these questions in the context of a model a spin--1/2 Kagome 
bilayer, which provides an analogue to the $\mathcal{Z}_2$ topological insulators 
considered by Kane and Mele \cite{Kane2005,Kane2005b,Fu2007}.  
We take a limit in which the ground state is a quantum paramagnet, formed by 
inter--layer dimers, with nine distinct bands of triplet excitations.
We consider first the case where the spin of triplet excitations is conserved, and show 
that in zero magnetic field the $S^z=\pm1$ triplet bands can realize an analogue to the 
quantum spin Hall insulator~\cite{Kane2005,Fu2007,Moore2007,Roy2009}.
In this case, triplon bands are characterized by a nonzero $\mathcal{Z}_{2}$ invariant, 
and in open geometries we find corresponding helical triplet edge modes.
Furthermore, when the time reversal symmetry is broken by applied magnetic field, the system evolves into a Chern insulator~\cite{Haldane1988} characterized by chiral triplon edge modes appearing in a finite sample. 
We compute the spin Nernst, and thermal Hall responses marking the nontrivial topology in these phases.


We explore the consequences of the spin--mixing terms allowed by the symmetry 
of the lattice, and discuss their effect on the $\mathcal{Z}_2$ and Chern bands. 
Such spin--mixing terms are also present in the original model of Kane and Mele, 
in the form of the Rashba coupling~\cite{Kane2005,Kane2005b}. 
In that case, the ``up'' and ``down'' spin states of an electron form a Kramers pair, 
enforcing the twofold degeneracy of the bands at certain points in the Brillouin zone.
This guarantees the perturbative stability of spin--Hall state against 
small values of  Rashba coupling.
However such a protection is {\it not} guaranteed by time--reversal symmetry in the case 
of the triplons, where the representation of the time--reversal operator squares to one. 
We examine this difference closely, and find that even infinitesimal nematic interactions 
can eliminate the $\mathcal{Z}_2$ topological invariant, opening a gap to the associated 
helical edge modes.


We further identify an operator $\Theta$=TR$\times$U(1) which, 
within a Bogoliubov-de Gennes approach, encodes the symmetry 
needed to protect a $\mathcal{Z}_2$ topological phase in either a quantum 
paramagnet, or an ordered phase with topological magnon bands.
We give a detailed analysis on the commutation of the various terms in the triplet 
Hamiltonian with $\Theta$, confirming that the nematic interaction breaks this symmetry, 
compromising the $\mathcal{Z}_2$ band-topology. 

We also discuss the implications of these results for the bilayer models 
considered in Refs.~\onlinecite{Joshi2019} and~\onlinecite{Kondo2019}, 
where identical symmetry considerations apply.


The remainder of the Article is organized as follows. 
In Section~\ref{sec:symm_Hamilton} we give a detailed analysis on the symmetry allowed intra-dimer and first neighbor inter-dimer interactions and introduce the bilayer kagome model.  \Secref{modelHamiltonian} is devoted to the Bogoliubov--de Gennes Hamiltonian describing the triplet dynamics, and provides a detailed discussion on the band touching topological transitions appearing as the anisotropies change.
In \Secref{tripletZ2TopologicalInsulator} we show that the triplet excitations can provide an analog to the Kane and Mele model~\cite{Kane2005}, characterized by nonzero $\mathcal{Z}_2$ topological invariant. We calculate the Nernst effect of triplets, a transverse spin current arising as a response to a temperature gradient. Sec.~\ref{sec:bad_nematic!} is devoted to the protection of the $\mathcal{Z}_2$ topology. We show that the TR$\times$U(1) symmetry corresponds to a pseudo time reversal operation, which, if present, protects the $\mathcal{Z}_2$ bands. However, symmetric exchange anisotropies, arising in the form of nematic interactions, break this symmetry, together with the fragile $\mathcal{Z}_2$ phase of a non-Kramers pairs. In Sec.~\ref{sec:tripletChernInsulator} we consider the time-reversal symmetry breaking case in the presence of magnetic field. We examine the stability of the triplet
Chern bands. Additionally, we compute the thermal Hall signal of the Chern-ful triplon bands.
\Secref{conclusions} provides a brief summary of our results.

\section{Symmetry-allowed Hamiltonian}
\label{sec:symm_Hamilton}
\begin{figure}[ht!]
	\begin{center}
		\includegraphics[width=1\columnwidth]{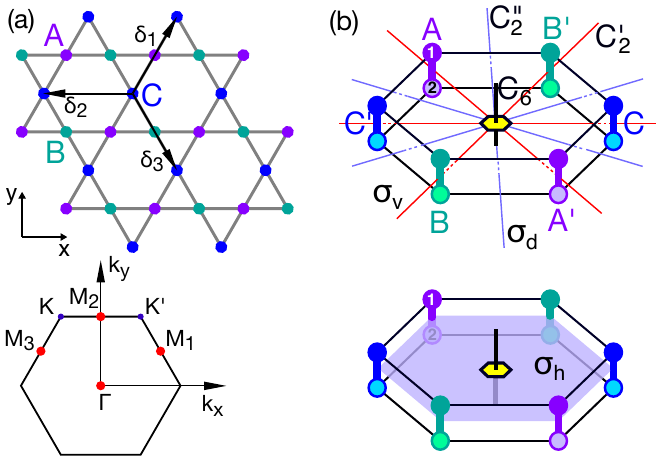}
		\caption{(a) Top view of the bilayer kagome lattice with the translation vectors, $\boldsymbol{\delta}_1=\left(1/2,\sqrt{3}/2\right)$, $\boldsymbol{\delta}_2=\left(-1, 0\right)$, and $\boldsymbol{\delta}_3=\left(1/2,-\sqrt{3}/2\right)$. Bottom panel: Hexagonal Brillouin zone with time reversal invariant momenta (TRIM) $\Gamma$, $\mathsf{M_1}$, $\mathsf{M_2}$, and $\mathsf{M_3}$ (red), as well as the Dirac points K, and K' (blue). (b) Symmetry operations of the bilayer-kagome lattice. The smallest unit exhibiting the full symmetry of the bilayer kagome lattice is a bilayer hexagon with six inter--layer dimers. The indices $A$, $B$, and $C$ and following the notation of the main text corresponding to the sublattice flavor, while the $A'$ denotes the dimer $A$ shifted with the lattice translation vector $\boldsymbol{\delta}_3$, $B'$ is dimer $B$ shifted with $\boldsymbol{\delta}_1$ and $C'$ corresponds to dimer $C$ translated by $\boldsymbol{\delta}_2$. }
		\label{fig:BZ_and_deltas}
	\end{center}
\end{figure}

The model we consider is the spin--1/2 magnet on the bilayer Kagome lattice, shown in Fig.~\ref{fig:BZ_and_deltas}(a).
We first establish the most general form of first--neighbour and second--neighbour 
interactions allowed by the D$_{6h}$ symmetry of this lattice.
The resulting model,  
\begin{equation}
   \mathcal{H}^{\text{1,2}}_{\text{D$_{6h}$}} = \mathcal{H}_{\text{XXZ}} + \mathcal{H}_{\text{DM}} + \mathcal{H}_{\text{Nematic}}  \; ,
   \label{eq:H.D6h}
\end{equation}
has fifteen adjustable parameters, and contains terms which we can group as symmetric XXZ exchanges, $\mathcal{H}_{\text{XXZ}}$; antisymmetric Dzyaloshinskii-Moriya (DM) interactions, $\mathcal{H}_{\text{DM}}$; and symmetric diagonal and off-diagonal exchange anisotropies, which we name the "bond--nematic" interactions (defined below), $\mathcal{H}_{\text{Nematic}}$. 

To determine the different contributions to Eq.~(\ref{eq:H.D6h}), we consider the hexagonal prism formed by six interlayer spin--dimers [Fig.~\ref{fig:BZ_and_deltas}(b)], which forms the smallest building block with the full symmetry of lattice.
In what follows, we refer to these interlayer dimers simply as ``dimers'', and we will ultimately build topological bands from the triplet excitations of a quantum paramgnet formed by singlets on these dimer bonds.
The bilayer hexagon contains three such dimers  $A$, $B$, and $C$ that lie within the kagome unit cell, as well as three more, $A'$, $B'$, and $C'$, which correspond to neighboring dimers translated by $\boldsymbol{\delta}_3$, $\boldsymbol{\delta}_1$, and $\boldsymbol{\delta}_2$, respectively [Fig.~\ref{fig:BZ_and_deltas}(a)].

We proceed by analysing all possible interactions bilinear in spins, starting from the transformation properties of individual spin components under three generators of the D$_{6h}$ point group, C$_6$, C$'_2$, and $\sigma_h$ [Table~\ref{tab:D6h}].
All remaining group elements can be constructed as a combinations of these three operations.
The different types of term which arise are considered, bond by bond, below.

\begin{table}[htp]
\caption{Transformation of the spin (axial vector) components, the dimers, and the site indices under the generators of the  D$_{6h}$ point group.}
\label{tab:D6h}
\begin{center}
\begin{ruledtabular}
\begin{tabular}{clccc}
 \multirow{1}{*}{\rotatebox[origin=c]{0}{\parbox[c]{1.6 cm}{\centering Generators}}} & E & C$_6$ & C$'_2$ & $\sigma_h$\\[0.5ex]
\hline
\multirow{3}{*}{\rotatebox[origin=c]{0}{\parbox[c]{1.6  cm}{\centering spin component}}} & $S^x$  & $\frac{1}{2} S^x+\frac{\sqrt{3}}{2}S^y$ & $S^x$ & $-S^x$\\
& $S^y$  & $-\frac{\sqrt{3}}{2} S^x+\frac{1}{2}S^y$ & $-S^y$ & $-S^y$\\
& $S^z$  & $S^z$ & $-S^z$ & $S^z$\\[0.5ex]
\hline
\multirow{6}{*}{\rotatebox[origin=c]{0}{\parbox[c]{1.6  cm}{\centering dimer label}}}  & $A$  & $C'$ & $B$ & $A$\\
& $B$ &$A'$ & $A$ & $B$\\
& $C$ & $B'$ & $C$ & $C$\\
& $A'$ & $C$ & $B'$ & $A'$\\
& $B'$ & $A$ & $A'$ & $B'$\\
& $C'$ & $B$ & $C'$ & $C'$\\[0.5ex]
\hline
\multirow{2}{*}{\rotatebox[origin=c]{0}{\parbox[c]{1.6  cm}{\centering layer index}}} & $1$   &  $1$  &  $2$  & $2$\\[0.5ex]
 & $2$   &  $2$  &  $1$  & $1$
\end{tabular}
\end{ruledtabular}
\end{center}
\label{default}
\end{table}%

\subsection{Intra-dimer interactions}\label{sec:intra-dimer}
\begin{figure}[h!]
	\begin{center}
		\includegraphics[width=0.8\columnwidth]{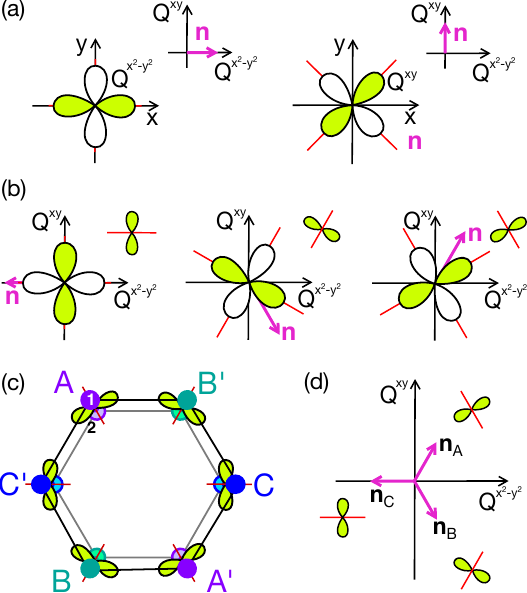}
		\caption{(a) Illustration of the components, $Q^{x^2\!-\!y^2}$ and $Q^{xy}$, of the intra-dimer nematic interactions. The pink arrows indicates the direction in the 2-dimensional vector-spaces, $\mathbf{Q}^{\|}=(Q^{x^2\!-\!y^2},Q^{xy})$. (b) The linear combinations appearing in the intra-dimer in-plane nematic terms. We plot the vectors $n$ and the schematic representation of the operators in the same figure. The sketches in the top right corner show the simplified illustration for a given linear combination. (c) Intra-dimer nematic operators on the bonds, and (d) the vectors $\mathbf{n}_A$, $\mathbf{n}_B$, and $\mathbf{n}_C$, corresponding to the linear combination on each dimer.}
		\label{fig:bond_nematic}
	\end{center}
\end{figure}

The symmetry classification of the intra-dimer interactions according to the D$_{6h}$ symmetry group yields three invariant terms. Two correspond to the Heisenberg exchange anisotropy distinguishing the in-plane and out-of-plane components

\begin{equation}
\mathcal{H}^{\singAB}_{\rm XXZ}=J_{\|}\sum_{j}(S^x_{j_1} S^x_{j_2}+S^y_{j_1} S^y_{j_2})+J_{\bot}\sum_{j}S^z_{j_1} S^z_{j_2}\;.
\label{eq:Heis_bond}
\end{equation}
The third intra-dimer term is a symmetric exchange anisotropy, which we refer to as the bond-nematic interaction:
\begin{equation}
\mathcal{H}^{\singAB}_{\rm Nematic}=K_{\|}\sum_{j}\mathbf{n}_j\cdot \mathbf{Q}^{\|}_{j_1,j_2}\;.
\label{eq:bond_nematic}
\end{equation}
The index $j$ runs over the dimers, and $1$ and $2$ denote the layer indices of dimer-$j$. The vectors $\mathbf{n}_j$ appearing in the nematic term have the form $\mathbf{n}_A=\left(\frac{1}{2},\frac{\sqrt{3}}{2}\right)$, $\mathbf{n}_B=\left(\frac{1}{2},-\frac{\sqrt{3}}{2}\right)$, and $\mathbf{n}_C=\left(-1,0\right)$, and $\mathbf{Q}^{\|}_{j_1,j_2}$ denotes the vector $(Q^{x^2\!-\!y^2}_{j_1,j_2},Q^{xy}_{j_1,j_2})$ made of the nematic interactions 
\begin{subequations}
\begin{eqnarray}
&Q^{x^2\!-\!y^2}_{i,j}=S^x_{i}S^x_{j}-S^y_{i}S^y_{j}\;,\\
&Q^{xy}_{i,j}=S^x_{i}S^y_{j}+S^y_{i}S^x_{j}\;.
\end{eqnarray}
\label{eq:in-plane-nematic}
\end{subequations}
We illustrate the nematic operators, $Q^{x^2\!-\!y^2}$ and $Q^{xy}$, in the fashion of the d-orbitals, $d^{x^2\!-\!y^2}$ and $d^{xy}$, as they transform in the same way. Fig.~\ref{fig:bond_nematic} introduces our schematic illustration of the in-plane nematic operators, $Q^{x^2\!-\!y^2}$ and $Q^{xy}$, and their linear combinations as appear in the intra-dimer interactions, together with the directions of the $\mathbf{n}$ vectors.

Due to the bond-inversion, the antisymmetric exchange anisotropy, i.e. the Dzyaloshinskii--Moriya interaction is not allowed on the dimers.

\subsection{First neighbor inter-dimer interactions}\label{sec:first_nb}
\begin{figure}[h!]
	\begin{center}
		\includegraphics[width=1\columnwidth]{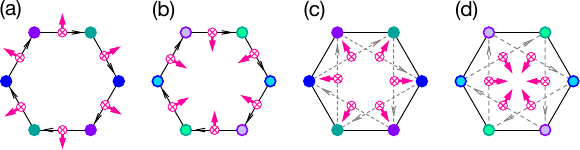}
		\caption{(a-b) DM vector components on the first, and (c-d) on the second neighbor inter-dimer bonds in the top and bottom layers, respectively. For both, the first and second neighbor DM interactions the in-plane components have opposite signs in the top and bottom layer, while the out-of-plane components are the same.}
		\label{fig:DM_vectors}
	\end{center}
\end{figure}

Classifying the first neighbor inter-dimer interactions, we find six independent operators that transform as the fully symmetric irreducible representation. Beside the anisotropy in the Heisenberg exchange, distinguishing the in-plane and out-of-plane components
\begin{eqnarray}
\mathcal{H}^{\rm 1st}_{\rm XXZ} &=&
J'_{\|}\sum_{\substack{\langle i,j \rangle \\ l=1,2}}(S^x_{i_l} S^x_{j_l}+S^y_{i_l} S^y_{j_l})+J'_{\bot}\sum_{\substack{\langle i,j \rangle \\ l=1,2}}S^z_{i_l} S^z_{j_l},
\label{eq:Heis_first}
\end{eqnarray}
there are additional four operators, namely the in-plane and out-of-plane components of the nematic and DM interactions. 
The DM interaction has the form 
\begin{eqnarray}
\mathcal{H}^{\rm 1st}_{\text{DM}} &=& 
 \sum_{\langle i,j\rangle} \mathbf{D}'\cdot(\mathbf{S}_i\times\mathbf{S}_j)\;,
\label{eq:H_DM}
\end{eqnarray}
and the in-plane ($D'_{\|}$) and out-of-plane ($D'_{\bot}$) components of the vector $\mathbf{D}'$ are shown in Fig.~\ref{fig:DM_vectors}. 

$D'_{\bot}$ is uniform in the top and bottom layers, but the $D'_{\|}$ components change sign under exchanging the layers. 
As discussed in Section~\ref{sec:modelHamiltonian}, only the uniform out-of-plane DM component $D'_{\bot}$ appears in Bogoliubov--de Gennes Hamiltonian describing the triplet dynamics.

\begin{figure}[h!]
	\begin{center}
		\includegraphics[width=1\columnwidth]{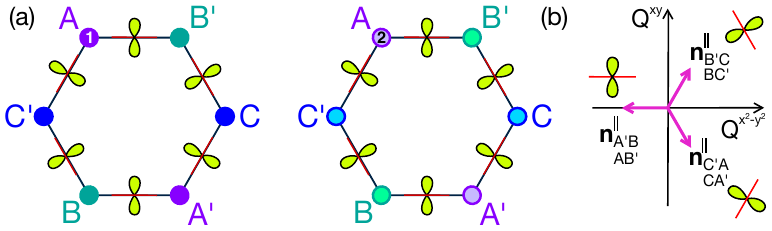}
		\caption{(a) Inter-dimer in-plane symmetric exchange anisotropy $\mathbf{Q}^{\|}_{i,j}$ in the top (left) and bottom (right) layers. (b) Components of the inter-dimer nematic operators between the first neighbor bonds.}
		\label{fig:1st_nb_in-plane_nematic}
	\end{center}
\end{figure}

In addition, there are in-plane and out-of-plane bond-nematic terms. The in-plane component has the form of
\begin{equation}
\mathcal{H}^{\rm 1st}_{ \|}=K'_{\|}\sum_{\langle i,j \rangle }\sum_{l=1,2}\mathbf{n}_{ij}\cdot \mathbf{Q}^{\|}_{il,jl}\;,
\label{eq:first_nematic}
\end{equation}
where $\mathbf{Q}^{\|}_{il,jl}$ is defined the same way as in Eq~\ref{eq:in-plane-nematic}, the index $l$ takes the value $1$ for the top, and $2$ for the bottom layer, while $i$ and $j$ denote first neighbor sites within the layers. 
The vectors $\mathbf{n}_{ij}$ have the form 
\begin{eqnarray}
\mathbf{n}_{A'B} &=& \mathbf{n}_{AB'} = \left(-1,0\right)  \; , \\ 
\mathbf{n}_{B'C} &= & \mathbf{n}_{BC'} = \left(\frac{1}{2},\frac{\sqrt{3}}{2}\right) \; , \\ 
\mathbf{n}_{C'A} &=& \mathbf{n}_{CA'} = \left(\frac{1}{2},-\frac{\sqrt{3}}{2}\right) \; .
\end{eqnarray} 
The in-plane component of the inter-dimer nematic interactions on the first neighbors are shown in Fig.~\ref{fig:1st_nb_in-plane_nematic}, where we use the same notation introduced in Fig.~\ref{fig:bond_nematic} (a) and (b). 

As with the out--of--plane DM components, the in--plane nematic terms are uniform in the top and bottom layers, and so will give a finite contribution to a Hamiltonian for triplets.

\begin{figure}[h!]
	\begin{center}
		\includegraphics[width=1\columnwidth]{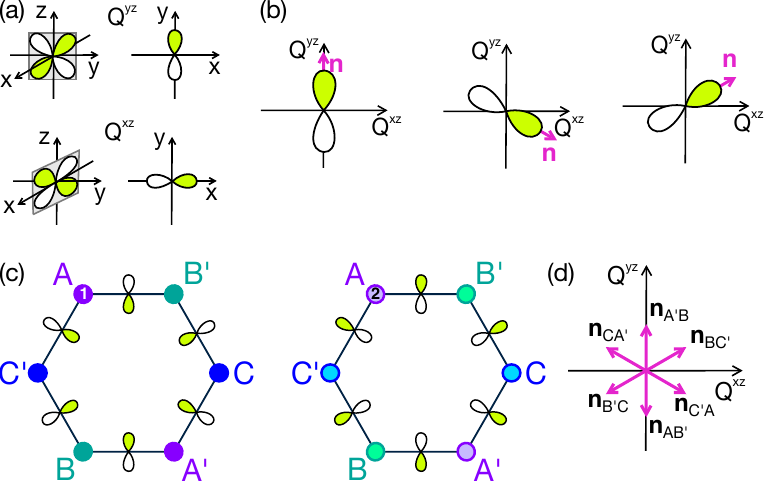}
		\caption{(a) Illustration of the out-of-plane nematic interaction components, $Q^{zx}$ and $Q^{yz}$.  The panels on the right show the top-view. (b) Some representative out-of-plane nematic terms and the corresponding  $\mathbf{n}^{\bot}_{ij}$ vectors (pink arrow), defining the linear combination of  $Q^{zx}$ and $Q^{yz}$. (c) Inter-dimer out-of-plane symmetric exchange anisotropy in the top (left) and bottom layer (right).  (d) Directions of the first neighbor out-of-plane nematic vectors in the top layer. The directions in the bottom layer are the opposite.}
		\label{fig:1st_nb_out-of-plane_nematic}
	\end{center}
\end{figure}

The first-neighbor out-of-plane bond-nematic interaction is
\begin{equation}
\mathcal{H}^{\rm 1st}_{\bot}=K'_{\bot}\sum_{\langle i,j \rangle }\mathbf{n}^{\bot}_{ij}\cdot \mathbf{Q}^{\bot}_{il,jl}\;,
\end{equation}
where the vectors $\mathbf{n}^{\bot}_{ij}$ are shown in Fig.~\ref{fig:1st_nb_out-of-plane_nematic}(b), and the components of $\mathbf{Q}^{\bot}_{i,j}=(Q^{zx}_{i,j},Q^{yz}_{i,j})$ correspond to the bond-nematic operators
\begin{subequations}
\begin{eqnarray}
&Q^{xz}_{i,j}=S^z_{i}S^x_{j}+S^x_{i}S^z_{j}\;,\\
&Q^{yz}_{i,j}=S^y_{i}S^z_{j}+S^z_{i}S^y_{j}\;.
\end{eqnarray}
\label{eq:out-of-plane-nematic}
\end{subequations} 
We utilize the representation of the $d^{yz}$ and $d^{zx}$ orbitals, to illustrate the out-of-plane bond-nematic terms as shown in Fig.~\ref{fig:1st_nb_out-of-plane_nematic}.

Once again, the out-of-plane inter-dimer symmetric exchange anisotropy term is the opposite in the top and bottom layers.
For this reason, as with the in-plane DM vectors, it will cancel in the triplet hopping Hamiltonian discussed in Section~\ref{sec:modelHamiltonian}.

\subsection{Second neighbor inter--dimer interactions}

We also consider the effect of second--neighbour interactions within the planes of the Kagome lattice.
As with the first--neighbor interactions, there are six different terms: 
XXZ exchange $J''_{\|}$ and $J''_{\bot}$; in--plane and out--of--plane DM interactions, $D''_{\|}$ and $D''_\bot$; and the in--plane and out--of--plane bond-nematic terms $K''_{\|}$ and $K''_{\bot}$. 
XXZ interactions are defined through
\begin{eqnarray}
\mathcal{H}^{\rm 2nd}_{\rm XXZ}&=&
J''_{\|}\sum_{\langle\langle i,j\rangle\rangle}\sum_{l=1,2}S^x_{i_l}  S^x_{j_l}+S^y_{i_l}  S^y_{j_l}\nonumber\\
&&+J''_{\bot}\sum_{\langle\langle i,j\rangle\rangle}\sum_{l=1,2}S^z_{i_l}   S^z_{j_l}\;,
\end{eqnarray}
where the sum $\langle\langle i,j \rangle\rangle$ runs over second--neighbor bonds within the Kagome layers.
Similarly, DM interactions are defined through
\begin{eqnarray}
\mathcal{H}^{\rm 2nd}_{\text{DM}} &=& 
 \sum_{\langle\langle i,j\rangle\rangle} \mathbf{D}'' \cdot(\mathbf{S}_i\times\mathbf{S}_j)\;,
\label{eq:H_DM_2}
\end{eqnarray}
where the components of the associated DM vectors in the top and bottom layers are shown in Fig.~\ref{fig:DM_vectors} (d) and (e), respectively.


The additional bond--nematic--type interactions have the form 
\begin{equation}
\mathcal{H}^{\rm 2nd}_{ \|}=K''_{\|}\sum_{\langle\langle i,j \rangle\rangle }\sum_{l=1,2}\mathbf{n}_{ij}\cdot \mathbf{Q}^{\|}_{il,jl}\;,
\label{eq:inplane_nematic_2}
\end{equation}
where the vector operator, $\mathbf{Q}^{\|}_{il,jl}$ is defined in Eq.~(\ref{eq:in-plane-nematic}); $l$ takes the values $1$ and $2$ for top and bottom layers, respectively and the vectors $\mathbf{n}_{ij}$ have the form 
\begin{eqnarray}
\mathbf{n}_{AB} &=& \mathbf{n}_{A'B'} = \left(1,0\right) \;, \\
\mathbf{n}_{BC} &=& \mathbf{n}_{B'C'} = \left(-\frac{1}{2},-\frac{\sqrt{3}}{2}\right) \;,  \\
\mathbf{n}_{CA} &=& \mathbf{n}_{C'A'} = \left(-\frac{1}{2},\frac{\sqrt{3}}{2}\right) \; .
\end{eqnarray}

\begin{figure}[h!]
	\begin{center}
		\includegraphics[width=1\columnwidth]{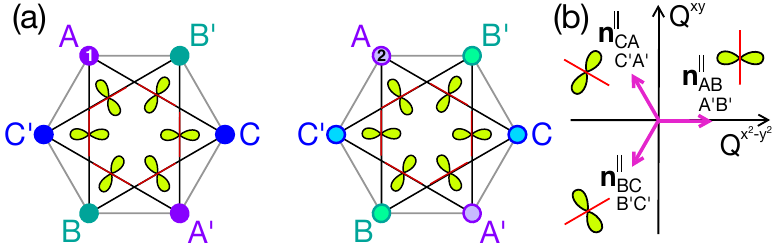}
		\caption{(a) Inter-dimer in-plane symmetric exchange anisotropy $\mathbf{Q}^{\|}_{i,j}$ between the second neighbors in the top (left) and bottom (right) layers. (b) Components of the inter-dimer nematic operators.}
		\label{fig:2nd_nb_in-plane_nematic}
	\end{center}
\end{figure}

\begin{figure}[h!]
	\begin{center}
		\includegraphics[width=1\columnwidth]{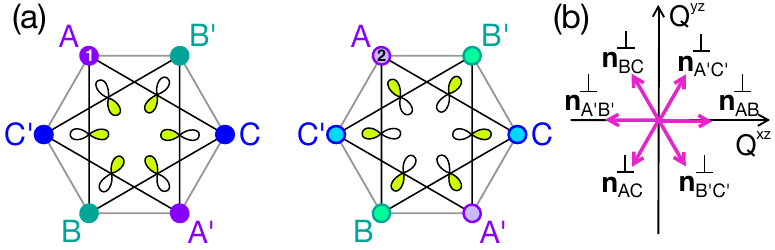}
		\caption{(a) Inter-dimer out-of-plane symmetric exchange anisotropy on the second neighbor bonds in the top (left) and bottom layer (right). We use the notation for the linear combinations of $Q^{zx}$ and $Q^{yz}$ bond-nematic operators as shown in Fig.~\ref{fig:1st_nb_out-of-plane_nematic}. (b) Directions of the second neighbor out-of-plane nematic vectors in the top layer. The directions in the bottom layer are the opposite.}
		\label{fig:2nd_nb_out-of-plane_nematic}
	\end{center}
\end{figure}

The out-of-plane bond-nematic term is
\begin{equation}
\mathcal{H}^{\rm 2nd}_{\bot}=K''_{\bot}\sum_{\langle\langle i,j \rangle\rangle }\sum_{l=1,2}\mathbf{n}^{\bot}_{ij}\cdot \mathbf{Q}^{\bot}_{il,jl}\;,
\end{equation}
where the components of $\mathbf{Q}^{\bot}_{i,j}=(Q^{zx}_{il,jl},Q^{yz}_{il,jl})$ are introduced in Eqs.~\ref{eq:out-of-plane-nematic} and the associated vectors $\mathbf{n}^{\bot}_{ij}$ are shown in Fig.~\ref{fig:2nd_nb_out-of-plane_nematic}(b). 

%

\section{Model of triplon bands}
\label{sec:modelHamiltonian}

Having established the most general form of interactions allowed by symmetry, we now show how this determines the Hamiltonian for triplet excitations of a quantum paramagnet on the bilayer Kagome lattice.
We begin by defining the model [Section~\ref{sec:spin.Hamiltonian}], before proceeding to a 
Hamiltonian expressed in terms of a bond--wave formalism, with 
explicit Bogoliubov--de Gennes form [Section~\ref{sec:bond.wave.Hamiltonian}].
Finally, we set up the framework for subsequent discussion of topological bands, 
by solving for the (topologically--trivial) triplon bands in the limit 
of vanishing anisotropic exchange [Section~\ref{sec:spin.conserving.limit}].
Throughout this Article, we work with the full Bogoliubov--de Gennes Hamiltonian for triplet 
excitations, except where otherwise stated.

\subsection{Microscopic Model}
\label{sec:spin.Hamiltonian}

We start from a model defined by
\begin{eqnarray}
   \mathcal{H} &=& \mathcal{H}^{\text{1,2}}_{\text{D$_{6h}$}} +  \mathcal{H}_{\text{Zeeman}} \\
   &=& \mathcal{H}_{\text{XXZ}} + \mathcal{H}_{\text{DM}} + \mathcal{H}_{\text{Nematic}} +  \mathcal{H}_{\text{Zeeman}}\; ,
   \label{eq:H}
\end{eqnarray}
where $ \mathcal{H}^{\text{1,2}}_{\text{D$_{6h}$}}$ [Eq.~(\ref{eq:H.D6h})] is the 
most general Hamiltonian for a spin--1/2 magnet on a bilayer Kagome lattice
with first-- and second--neighbour interactions [cf. Section~\ref{sec:symm_Hamilton}], 
and 
\begin{eqnarray}
\mathcal{H}_{\text{Zeeman}} = -g_z h^z \sum_i S^z_i \; ,
\label{eq:H.Zeeman}
\end{eqnarray}
encodes the effect of a magnetic field perpendicular to the 
plane of the Kagome lattice.


The largest term in this model is taken to be (approximately) Heisenberg
exchange $J_\parallel \approx J_\perp$ on intra--dimer bonds [Section~\ref{sec:intra-dimer}]. 
Where these interactions are antiferromagnetic, and sufficiently large compared with other terms, the ground state of $\mathcal{H}_{\text{XXZ}}$ is a quantum paramagnet formed by a product of singlets on all dimer bonds 
 \begin{eqnarray}
 	\mid \Psi_0 \rangle = \prod_{j\in\singAB}|s\rangle_j \; , 
 \end{eqnarray}
where
  \begin{eqnarray}
 \left|s\right>_j=\frac{1}{\sqrt{2}}(\left|\uparrow_1\downarrow_2\right>-\left|\downarrow_1\uparrow_2\right>)_j \; .
 \end{eqnarray} 


The low--lying excitations of this quantum paramagnet will be spin-1 triplet excitations, 
\begin{subequations} 
\begin{eqnarray}
&&  \left|t_1\right>_j = \left|\uparrow_1\uparrow_2\right>_j \\
&&  \left|t_0\right>_j = \frac{1}{\sqrt{2}}(\left|\uparrow_1\downarrow_2\right>+\left|\downarrow_1\uparrow_2\right)_j \\
&&  \left|t_{-1}\right>_j= \left|\downarrow_1\downarrow_2\right>_j \;,
 \end{eqnarray}
   \label{eq:tn}
 \end{subequations} 
where the indices $1$ and $2$ denote the top and bottom sites of the dimer, and $j$ labels the dimer. 
In the presence of terms connecting different dimers, these triplet excitations 
will form dispersing bands of excitations, usually referred to as ``triplons''.

The spin operators can be expressed using the singlet-triplet basis
\begin{subequations} 
 \begin{eqnarray}
 S^+_{j_1} &\!\!=\!\!& 
 \frac{1}{\sqrt{2}} \left(
 t^\dagger_{1,j} t^{\phantom{\dagger}}_{0,j} 
 \!- \!  t^\dagger_{1,j}  s^{\phantom{\dagger}}_{j}
 +
 t^\dagger_{0,j}t^{\phantom{\dagger}}_{-1,j}
 \!+\! s^\dagger_{j}t^{\phantom{\dagger}}_{-1,j}
 \right)
 \\ 
 S^z_{j_1} &\!\!=\!\!& \frac{1}{2}\left(
 t^\dagger_{1,j} t^{\phantom{\dagger}}_{1,j}
 + t^\dagger_{0,j} s^{\phantom{\dagger}}_{j} 
+ s^\dagger_{j} t^{\phantom{\dagger}}_{0,j} 
 - t^\dagger_{-1,j} t^{\phantom{\dagger}}_{-1,j}
 \right)
 \\ 
 S^-_{j_1} &\!\!=\!\!& 
 \frac{1}{\sqrt{2}} \left(
t^\dagger_{0,j}  t^{\phantom{\dagger}}_{1,j}
 \!-\! s^\dagger_{j} t^{\phantom{\dagger}}_{1,j}
 +
 t^\dagger_{-1,j} t^{\phantom{\dagger}}_{0,j} 
 \!+\!  t^\dagger_{-1,j} s^{\phantom{\dagger}}_{j} 
 \right)
 \\ 
 S^+_{j_2} &\!\!=\!\!& 
 \frac{1}{\sqrt{2}} \left(
 t^\dagger_{1,j} t^{\phantom{\dagger}}_{0,j} 
 \!+\!  t^\dagger_{1,j} s^{\phantom{\dagger}}_{j} 
 +
 t^\dagger_{0,j} t^{\phantom{\dagger}}_{-1,j}
 \!- \!s^\dagger_{j} t^{\phantom{\dagger}}_{-1,j}
 \right) 
 \\ 
 S^z_{j_2} &\!\!=\!\!& \frac{1}{2}\left(
 t^\dagger_{1,j} t^{\phantom{\dagger}}_{1,j}
 - t^\dagger_{0,j} s^{\phantom{\dagger}}_{j} 
- s^\dagger_{j} t^{\phantom{\dagger}}_{0,j} 
 - t^\dagger_{-1,j} t^{\phantom{\dagger}}_{-1,j}
 \right)
 \\ 
 S^-_{j_2} &\!\!=\!\!& 
 \frac{1}{\sqrt{2}} \left(
 t^\dagger_{0,j} t^{\phantom{\dagger}}_{1,j}
\!+\! s^\dagger_{j}t^{\phantom{\dagger}}_{1,j} 
 +
 t^\dagger_{-1,j} t^{\phantom{\dagger}}_{0,j} 
 \!-\!  t^\dagger_{-1,j} s^{\phantom{\dagger}}_{j} 
 \right) \end{eqnarray}
 \label{eq:tn}
 \end{subequations} 
where the indices $1$ and $2$ denote the top and bottom site (layer) of the vertical dimer, $t^\dagger_{m,j}\left|0\right>=\left|t_{m}\right>_j$, $s^\dagger_{j}\left|0\right>=\left|s\right>_j$ with $m=-1,0,1$ and the $x$ and $y$ components of the spin operators are $S^x_j=\frac 1 2(S^+_j+S^-_j)$ and $S^y_j=-\frac i 2(S^+_j-S^-_j)$.


In general, all components of the different exchange interactions within the Kagome 
planes will contribute to triplon dispersion.
For small to moderate spin--orbit coupling, the largest contributions will come from 
(approximately) Heisenberg interactions on first--neighbour bonds $J'_\parallel  \approx J'_\perp$.
However the Berry phase underpinning a topological bandstructure originates 
in the DM interactions, $\mathcal{H}_{\text{DM}}$.
And the fate of these topological bands will in turn depend on the 
nematic interactions, $\mathcal{H}_{\text{Nematic}}$.
A few comments are therefore due on how these enter the problem.


Our model allows for DM interactions on first--neighbour bonds, $\mathbf{D}'$ [Eq.~(\ref{eq:H_DM})], and second--neighbour bonds, $\mathbf{D}''$ [Eq.~(\ref{eq:H_DM_2})], illustrated in Fig.~\ref{fig:DM_vectors}. 
The inter-layer bond--inversion symmetry precludes any DM interaction acting 
on the local dimers, and consequently the mixing between the odd singlet and 
even triplet states.
Furthermore, as long as the $\sigma_h$ mirror symmetry is preserved, only the uniform out-of-plane DM components, 
$\mathbf{D}' = (0,0,D')$ and $\mathbf{D}'' = (0,0,D'')$, survive 
in the triplet Hamiltonian to linear order.  


Similar considerations apply to nematic interactions $\mathcal{H}_{\text{Nematic}}$.
In this case, the sign of the out-of-plane nematic terms is opposite in the layers, leaving only the in-plane components in the triplet Hamiltonian.
$\mathcal{H}_{\text{Nematic}}$ thus simplifies to only the intra--dimer and inter--dimer in--plane nematic interactions, introduced in Eq.~(\ref{eq:bond_nematic}), Eq.~(\ref{eq:first_nematic}), and~(\ref{eq:inplane_nematic_2}).


Without loss of generality, we consider only the first neighbour interactions. The further neighbor terms do not change the overall shape (and generality) of the triplet Hamiltonian, only add to the complexity of the coefficients. Though, we keep the second neighbour DM interaction ($D''$) to discuss band touching topological transitions as the functions of $D'$ and  $D''$.

In the remainder of this paper we will slowly build up a complete picture 
of the topological physics of a quantum paramagnet described by Eq.~(\ref{eq:H}), 
starting in Section~\ref{sec:band_topo} from a simplified model with only Heisenberg and 
DM interactions, before progressively restoring the complexity of the full Hamiltonian, 
including nematic terms. %
Before doing so, in what follows, we set up the necessary technical framework for evaluating triplon bands.

\subsection{Bond-wave Hamiltonian}\label{sec:bond-wave}
\label{sec:bond.wave.Hamiltonian}

To describe the dynamics of the triplet excitations at momentum  ${\bf k}$ , we rely on the conventional 
bond wave theory\cite{Sachdev1990,Collins2006} resulting in the Bogoliubov--de Gennes Hamiltonian
\begin{equation}
\mathcal{H}_k = 
\begin{pmatrix}
 \tilde{\mathbf{t}}^{\dagger}_{\mathbf{k}}\\
\tilde{\mathbf{t}}^{\phantom{\dagger}}_{-\!\bf k}
\end{pmatrix}^T
\begin{pmatrix}
\tilde M_{\mathbf{k}}& \tilde N_{\mathbf{k}}\\
\tilde N^\dagger_{\mathbf{-k}} & \tilde M^\dagger_{\mathbf{-k}}
\end{pmatrix}
\begin{pmatrix}
\tilde{\mathbf{t}}^{\phantom{\dagger}}_{\mathbf{k}}\\
\tilde{\mathbf{t}}^{\dagger}_{-\!\bf k}
\end{pmatrix}.
\label{eq:triplet_BdG_18x18}
\end{equation}
The vector $\tilde{\mathbf{t}}^{\dagger}_\mathbf{k}$ contains the 9 different triplets excitations, corresponding to the three spin states and the three sublattices; 
\begin{equation}
\tilde{\mathbf{t}}^{\dagger}_{\mathbf{k}}
  = \left(
\mathbf{t}_{1,{\bf k}}^{\dagger},\mathbf{t}_{0,{\bf k}}^{\dagger},\mathbf{t}_{-1,{\bf k}}^{\dagger} \right),
\label{eq:tdmkdef}
\end{equation}
where $\mathbf{t}_{m,{\bf k}}^{\dagger}$ ($m=-1,0,1$) is
\begin{equation}
\mathbf{t}_{m,{\bf k}}^{\dagger} = 
 (t_{A,m,{\bf k}}^{\dagger}, t_{B,m,{\bf k}}^{\dagger}, t_{C,m,{\bf k}}^{\dagger}).
\end{equation} 
 $M_{\mathbf{k}}$ and $N_{\mathbf{k}}$ are 9-by-9 matrices, corresponding to the hopping hamiltonian, and the pairing terms, respectively. As it will turn out below, the matrices $\tilde M_{\mathbf{k}}$ and $\tilde N_{\mathbf{k}}$ are Hermitian and contain only cosines of the wave vector, so they are even in $\mathbf{k}$: $\tilde M^{\phantom{\dagger}}_{\mathbf{k}}= \tilde M^{\dagger}_{\mathbf{k}} = \tilde M^{\phantom{\dagger}}_{-\mathbf{k}}$ and $\tilde N^{\phantom{\dagger}}_{\mathbf{k}} = \tilde N^{\dagger}_{\mathbf{k}} = \tilde N^{\phantom{\dagger}}_{-\mathbf{k}}$. This will simplify the formulas, also the action of the time-reversal operator.

In the presence of the  $\sigma_h$ mirror plane, the  $\left|t_0\right>$ triplet decouples from the time-reversal pair $\left|t_1\right>$ and $\left|t_{-1}\right>$ 
and the most general form of the Bogoliubov--de Gennes Hamiltonian, describing the triplet dynamics, decomposes into two blocks,
\begin{subequations}
\begin{align}
\mathcal{H}_\mathbf{k}^{(1,-1)} &= 
\begin{pmatrix}
\mathbf{t}_{{\bf k}}^{\dagger}\\
\mathbf{t}_{-\!{\bf k}}^{\phantom{\dagger}}
\end{pmatrix}^T
\begin{pmatrix}
M_{\mathbf{k}}  \!+\! M^{\text{Zeeman}}_{\mathbf{k}} & N_{\mathbf{k}}\\
N_{\mathbf{k}} & M_{\mathbf{k}} \!-\! M^{\text{Zeeman}}_{\mathbf{k}} 
\end{pmatrix}
\begin{pmatrix}
\mathbf{t}_{{\bf k}}^{\phantom{\dagger}}\\
\mathbf{t}_{-\!{\bf k}}^{\dagger}
\end{pmatrix}
\label{eq:triplet_BdG}
\\
\mathcal{H}_\mathbf{k}^{(0)} & = 
\begin{pmatrix}
\mathbf{t}_{0,{\bf k}}^{\dagger}\\
\mathbf{t}_{0,-\!{\bf k}}^{\phantom{\dagger}}
\end{pmatrix}^T
\begin{pmatrix}
M_{0,\mathbf{k}}& N_{0,\mathbf{k}}\\
N_{0,\mathbf{k}} & M_{0,\mathbf{k}} 
\end{pmatrix}
\begin{pmatrix}
\mathbf{t}_{0,{\bf k}}^{\phantom{\dagger}}\\
\mathbf{t}_{0,-\!{\bf k}}^{\dagger}
\end{pmatrix}
\label{eq:triplet0_BdG}
\end{align}
\end{subequations}
where the $M_{\mathbf{k}}$ and $N_{\mathbf{k}}$ are 6-by-6, and $M_{0,\mathbf{k}}$, and  $N_{0,\mathbf{k}}$ are 3-by-3 matrices. 
The spinful subspace in Eq.~(\ref{eq:triplet_BdG}) is spanned by
%
%
\begin{equation}
\mathbf{t}_{{\bf k}}^{\phantom{\dagger}} =
\begin{pmatrix}
 \mathbf{t}_{1,{\bf k}}^{\phantom{\dagger}}\\
  \mathbf{t}_{-1,{\bf k}}^{\phantom{\dagger}}
  \end{pmatrix} \quad\text{and}\quad 
\mathbf{t}_{-{\bf k}}^\dagger=
\begin{pmatrix}
 \mathbf{t}_{-1,-{\bf k}}^\dagger \\ 
 \mathbf{t}_{1,-{\bf k}}^\dagger 
\end{pmatrix} .
\label{eq:basis}
\end{equation}
Using this basis, both the diagonal and off-diagonal matrices are Hermitian, 
$M^\dagger_{\mathbf{k}}=M^{\phantom\dagger}_{\mathbf{k}}$, 
$M^\dagger_{0,\mathbf{k}}=M^{\phantom\dagger}_{0,\mathbf{k}}$, 
$N^\dagger_{\mathbf{k}}=N^{\phantom\dagger}_{\mathbf{k}}$,  
 and $N^\dagger_{0,\mathbf{k}}=N^{\phantom{\dagger}}_{0,\mathbf{k}}$. 
 The $\sigma_h$ acts on the individual $S=1/2$ spins as $C_2 = C_6^3$ combined with swapping the layer indices. As a consequence, the  $\left|t_0\right>$ tranforms differently from the $\left|t_{\pm1}\right>$ and $\left|s\right>$ under the   $\sigma_h$ operation: while  $\left|t_0\right> \to -\left|t_0\right>$, the $\left|t_{\pm1}\right> \to \left|t_{\pm1}\right>$  and $\left|s\right> \to \left|s\right>$. This happens for example in the terms containing  $S^x_i S^z_j$ or $S^y_i S^z_j$. Such terms are present in the in-plane DM and in the out-of-plane nematic interactions, resulting in  $t^{(\dagger)}_{0,i} t^{(\dagger)}_{\pm 1,j}$ and  $t^{(\dagger)}_{\pm 1, i} t^{(\dagger)}_{0,j}$ terms. These terms are odd under the $\sigma_h$ reflection, and therefore cancel in the triplet Hamiltonian 
\footnote{Let us mention, however, that a magnetic field applied in the $xy$-plane induces mixing between the spinful ($\left|t_1\right>$ and $\left|t_{-1}\right>$) and the spinless ($\left|t_0\right>$) subspaces.}. 
Due to the cancellation of the in-plane DM terms in the Hamiltonian, the only terms that do not conserve the total $S^z_T$ are the the in-plane nematic interactions, which change the $S^z_T$ by $\pm2$ by creating an $\left|t_1\right>$ from $\left|t_{-1}\right>$ and vice versa.  
 
We use the 8 Gell-Mann matrices as the basis for the sublattice degrees of freedom, corresponding to the three dimers, $A$, $B$, and $C$ in the unit cell:
\begin{gather}
\lambda_1\! =\!
\!\!\left(\begin{array}{ccc}
0 & 1 & 0\\
1& 0 & 0\\
0 & 0 & 0
\end{array}\right)\!\!,
\lambda_2\! =\!
\!\!\left(\begin{array}{ccc}
0 & -i & 0\\
i & 0 & 0\\
0 & 0 & 0
\end{array}\right)\!\!,
\lambda_3=\!
\!\!\left(\begin{array}{ccc}
1 & 0 & 0\\
0& -1 & 0\\
0 & 0 & 0
\end{array}\right)\!\!,\nonumber\\
\lambda_4\! =\!
\!\!\left(\begin{array}{ccc}
0 &0 & 1\\
0& 0 & 0\\
1 & 0 & 0
\end{array}\right)\!\!,
\lambda_5\! =\!
\!\!\left(\begin{array}{ccc}
0 & 0& i\\
0& 0 & 0\\
-i & 0 & 0
\end{array}\right)\!\!,
\lambda_6=\!
\!\!\left(\begin{array}{ccc}
0 & 0 & 0\\
0& 0 & 1\\
0 & 1 & 0
\end{array}\right)\!\!,\nonumber\\
\lambda_7\! =\!
\!\!\left(\begin{array}{ccc}
0 &0 & 0\\
0& 0 & -i\\
0 & i & 0
\end{array}\right)\!\!,
\lambda_8\! =\! \frac{1}{\sqrt{3}}
\!\!\left(\begin{array}{ccc}
1 & 0& 0\\
0& 1& 0\\
0& 0 & -2
\end{array}\right)\!\!.
\end{gather}
The Gell-Mann matrices, and the 3-by-3 identity matrix, $I_3$ sufficiently characterize the spinless  $m=0$ subspace. 
The hopping Hamiltonian has the form
\begin{eqnarray}
M_{0,\mathbf{k}}=M^{\text{XXZ}}_{0,\mathbf{k}}&=&J_\| I_3 + J'_\bot \left[\cos\frac{\boldsymbol{\delta}_1\mathbf{k}}{2} \lambda_4\right. \nonumber\\
&&\left.+ \cos\frac{\boldsymbol{\delta}_2\mathbf{k}}{2} \lambda_1+ \cos\frac{\boldsymbol{\delta}_3\mathbf{k}}{2}\lambda_6\right]\;,
\end{eqnarray}
and the pairing terms are the same as $M_{0,\mathbf{k}}$ without the diagonal elements.
\begin{eqnarray}
N_{0,\mathbf{k}}=M_{0,\mathbf{k}}-J_\|  I_3\;.
\end{eqnarray}
Note that the Hamiltonian in the $m=0$ subspace contains only the Heisenberg terms, the symmetric nematic exchange and the antisymmetric DM interaction do not affect the $\left|t_0\right>$ triplet. 
 
The triplet hopping of the spinful subspace,
\begin{equation}
M_{\mathbf{k}} = M^{\text{XXZ}}_{\mathbf{k}} + M^{\text{DM}}_{\mathbf{k}} + M^{\text{Nematic}}_{\mathbf{k}} 
 \label{eq:Msigmak}
\end{equation}
in Eq.~(\ref{eq:triplet_BdG}) is a $6\times 6$ matrix.  The spin degree of freedom provided by the $S^z=\pm 1$ triplets is represented by the spin operators $s^x$, $s^y$, and $s^z$, corresponding to the Pauli matrices times $\frac 1 2$. The 6-dimensional local Hilbert space for the spinful triplets is constructed as the tensor product of the 2-by-2 matrices $\{I_2,s^x, s^y, s^z\}$ acting on the spin-space, and the 3-by-3 Gell-Mann matrices extended with the identity $I_3$, acting in the sublattice space. We discuss the various contributions separately.
The Heisenberg interaction only contains the identity operator, $I_2$ in the spin-space, i.e. it does not affect the spin degrees of freedom, and is the same for the $\left|t_1\right>$ and $\left|t_{-1}\right>$ triplets. 
\begin{eqnarray}
M^{\text{XXZ}}_{\mathbf{k}}&\!=\!&\frac{J_\|\!+\!J_\bot}{2} I_2\!\otimes\! I_3 + J'_\| \cos\frac{\boldsymbol{\delta}_1\mathbf{k}}{2} I_2\!\otimes \!\lambda_4\nonumber\\
&&+J'_\| \cos\frac{\boldsymbol{\delta}_2\mathbf{k}}{2} I_2\!\otimes \!\lambda_1+J'_\| \cos\frac{\boldsymbol{\delta}_3\mathbf{k}}{2} I_2
\!\otimes\! \lambda_6.
\label{eq:MXXZ}
\end{eqnarray}
The pairing terms from the Heisenberg interaction are similar to $M^{\text{XXZ}}_{\mathbf{k}}$, but have opposite sign and no diagonal elements
\begin{eqnarray}
 N^{\text{XXZ}}_{\mathbf{k}}&=&-M^{\text{XXZ}}_{\mathbf{k}}+\frac{J_\|+J\bot}{2} I_2\otimes I_3\;.
 \end{eqnarray}
 
The DM interaction has the form
\begin{eqnarray}
M^{\text{DM}}_{\mathbf{k}}&\!=\!&
\left[\!D' \! \cos\frac{\boldsymbol{\delta}_1\mathbf{k}}{2}\!+\!D'' \cos\!\frac{(\boldsymbol{\delta}_2\!-\!\boldsymbol{\delta}_3)\mathbf{k}}{2}\!\right]\! s^z \! \otimes \! \lambda_5\nonumber\\
&+&\left[\!D' \! \cos\frac{\boldsymbol{\delta}_2\mathbf{k}}{2}\!+\!D'' \cos\!\frac{(\boldsymbol{\delta}_3\!-\!\boldsymbol{\delta}_1)\mathbf{k}}{2}\!\right]\! s^z \! \otimes \! \lambda_2\nonumber\\
&+&\left[\!D' \! \cos\frac{\boldsymbol{\delta}_3\mathbf{k}}{2}\!+\!D'' \cos\!\frac{(\boldsymbol{\delta}_1\!-\!\boldsymbol{\delta}_2)\mathbf{k}}{2}\!\right]\! s^z \! \otimes \! \lambda_7\;.
\label{eq:DM_hopp}
\end{eqnarray}
The intra-dimer DM interaction is forbidden by the bond-inversion of the dimers, thus there are no diagonal elements in $M^{\text{DM}}_{\mathbf{k}}$
and the form of the pairing terms simply correspond to 
 \begin{eqnarray}
 N^{\text{DM}}_{\mathbf{k}}&=&-M^{\text{DM}}_{\mathbf{k}}\;.
 \end{eqnarray}
Here, the only operator acting in the spin-space is $s^z$, leaving the spin degrees of freedom unchanged, and introducing a sign difference for the DM interaction in the up and down-spin sector. 

Let us make some comments on the time-reversal (TR) properties of the Gell-Mann matrices and the pseudo-spin-half operators. The Gell-Mann matrices act in the sublattice space, i.e. account for changing the dimer indices, $A$, $B$, and $C$. Time-reversal symmetry leaves such indices invariant, therefore the real Gell-Mann matrices are TR invariant. As TR symmetry contains a complex conjugation, the imaginary Gell-Matrices are TR breaking, changing sign under TR. A complete analysis on the TR symmetry of the pseudo-spin-half operators is provided in the Appendix~\ref{app:TR_symm}, where we show that while the $s^z$ is TR-breaking, as one would expect from a spin-operator, the $s^x$ and $s^y$ components are TR invariant operators. This is a consequential difference with respect to the original Kane and Mele model, where the Pauli matrices describe a physical spin-half Kramers doublet and thus all break TR symmetry.  %

Coming back to the DM terms in our triplet hopping Hamiltonian in Eq.~(\ref{eq:DM_hopp}), the appearing cross-product operators are all TR invariant, as both the $s^z$ spin operator, and the $\lambda_5$, $\lambda_2$, and $\lambda_7$ complex Gell-Mann matrices are odd under TR symmetry.

The nematic interactions couple the subspaces of the different $m$ subspaces. They have the form
\begin{eqnarray}
M_{\mathbf{k}}^{\text{Nematic}}&\!=\!&
\frac{\sqrt{3} K_\| }{4}\left[s^x\otimes \lambda_8-s^y\otimes\lambda_3\right]\nonumber\\
&&+K'_\| \cos\frac{\boldsymbol{\delta}_3\mathbf{k}}{2} \left(\!\!-\frac{1}{2} s^x \!-\! \frac{\sqrt{3}}{2} s^y\! \!\right) \!\otimes\! \lambda_6\nonumber\\
&&+K'_\| \cos\frac{\boldsymbol{\delta}_1\mathbf{k}}{2} \left(\!\!-\frac{1}{2}  s^x\!+\! \frac{\sqrt{3}}{2} s^y\!\! \right)\!\otimes \!\lambda_4\nonumber\\
&&+K'_\| \cos\frac{\boldsymbol{\delta}_2\mathbf{k}}{2} \cdot s^x \!\otimes\!\lambda_1\;,
\label{eq:Mpm1_Nematic}
\end{eqnarray}
The Gell-Mann matrices in the nematic interaction are the real ones, preserving TR symmetry. The nematic interactions exclusively consist of spin-mixing operators, $s^x$ and $s^y$ that are TR invariant terms themselves. (For details see App.~\ref{app:TR_symm}). When the nematic terms are present, the total $S^z_T$ ceases to be a good quantum number, and the spin up and down components mix.

Such spin-mixing term is present in the Kane and Mele model too, in the form of a Rashba spin-orbit coupling, permitted by the breaking of the mirror-symmetry $\sigma_h$. Here, the spin-mixing nematic terms are allowed without breaking $\sigma_h$. The TR symmetry does not protect the degeneracy of the spin-up and down triplets at the TR-invariant points in the momentum space, as it would for the Kramers pair electron-spins. Therefore, the nematic term hybridizes the spins, ending the fragile $\mathcal{Z}_2$ topology of the bands, as discussed in Sec.~\ref{sec:bad_nematic!}.

The nematic interaction is allowed on the dimers too. Note that the operators $\lambda_8$ and $\lambda_3$ are diagonal in the sublattice space. The $s^x$ and $s^y$ operators, however mix the spins, placing the intra-dimer nematic interaction $K_\| $ in the diagonal of the block connecting the $+1$ and $-1$ triplets. The $N^{\text{Nematic}}_{1,\mathbf{k}}$ matrix of the pairing terms corresponds again to the $-M^{\text{Nematic}}_{1,\mathbf{k}}$ minus the `diagonal' elements
 \begin{eqnarray}
	N^{\text{Nematic}}_{\mathbf{k}}\!=\!-M^{\text{Nematic}}_{\mathbf{k}}\!+\!\frac{\sqrt{3} K_\| }{4}\left[s^x\!\otimes\! \lambda_8\!-\!s^y\!\otimes\!\lambda_3\right]\;.
 \end{eqnarray}

Lastly, the out-of plane magnetic field appears in the diagonal of the Hamiltonian as
\begin{eqnarray}
M_{\mathbf{k}}^{\text{Zeeman}}&\!=\!&
- g_z h_z 2 s^z\otimes I_3\;.
\label{eq:Mpm1_Zee}
\end{eqnarray}


We conclude this tour of the terms in the quadratic Bogoliubov--de Gennes Hamiltonian 
with a brief comment on what it neglects, namely interactions between 
triplon modes occurring at higher order in bond operators.
The effect of electron--electron interactions on topological insulators 
and superconductors remains an open problem.
And the effect of triplon--triplon interactions on band topology in quantum 
paramagnets is even less explored.

None the less, some work has been done the characterize the effect 
of magnon--magnon interactions in magnetically--ordered systems 
with topological bands \cite{Chernyshev2015,Chernyshev2016,McClarty2018,McClarty2019,Kondo2020,McClarty-arXiv}.
One example which has been been quite well characterized in the 
Kitaev model in high magnetic field, where the non--interacting  
theory of topological magnon bands can be compared with both 
interacting spin--wave calculations and DMRG results \cite{McClarty2018}.
And here the effect of interactions is relatively benign, 
being chiefly limited to a finite broadening of magnon modes, 
and renormalisation of band their dispersion.

Its is reasonable to expect the same will be true in quantum paramagnets,
since the form of Bogoliubov--de Gennes Hamiltonian is identical.
And this particulary true where the potential for triplon decay is restricted 
by a substantial band gap.
It is also plausible that the topological excitations of quantum 
paramagnets will exhibit some of the same interesting, non--Hermitian 
features, as topological magnons, a topic reviewed in \cite{McClarty-arXiv}.
The role of interactions within a $\mathcal{Z}_2$ topological phase of a magnetic insulator 
is clearly worthy of further investigation.
However this lies outside the scope of this Article, which 
has the more limited goal of characterizing band topology in the 
non--interacting limit.

\subsection{Bond-wave dispersions in the absence of anisotropies}
\label{sec:spin.conserving.limit}

Let us start with the time reversal symmetric case, when the magnetic field is zero. In the isotropic Heisenberg limit, $J_\bot=J_\|=J$, and $J'_\bot=J'_\|=J'$, $D'=0$, $D''=0$, $K_\|=0$, and $K'_\|=0$ the model has SU(2) symmetry, the $M_{\mathbf{k}}$ and $N_{\mathbf{k}}$ matrix in Eq.~(\ref{eq:triplet_BdG}) becomes block diagonal, 
\begin{equation}
M_{\mathbf{k}} = \begin{pmatrix}
M_{1,\mathbf{k}} & 0 \\
0 & M_{-1,\mathbf{k}}
\end{pmatrix}\;,
\quad
N_{\mathbf{k}} = \begin{pmatrix}
N_{1,\mathbf{k}} & 0 \\
0 & N_{-1,\mathbf{k}}
\end{pmatrix}\;,
\end{equation} 
and the Hamiltonians is identical in each of the $m=-1,0,1$ subspaces 
(see Eqs.~(\ref{eq:triplet_BdG})  and (\ref{eq:triplet0_BdG})) 
so that $M_{m,\mathbf{k}}=M^{\text{SU(2)}}_{m,\mathbf{k}}$, and  $N_{m,\mathbf{k}}=N^{\text{SU(2)}}_{m,\mathbf{k}}$, with $m$-independent 
\begin{eqnarray}
M^{\text{SU(2)}}_{m,\mathbf{k}}&\!\!=\!\!&J I_3 \!+\! J'\!\!\left[\cos\frac{\boldsymbol{\delta}_1\mathbf{k}}{2} \lambda_4\!+\! \cos\frac{\boldsymbol{\delta}_2\mathbf{k}}{2} \lambda_1\!+\!\cos\frac{\boldsymbol{\delta}_3\mathbf{k}}{2} \lambda_6\right]\!,\nonumber\\
N^{\text{SU(2)}}_{m,\mathbf{k}}&\!\!=\!\!&J I_3-M^{\text{SU(2)}}_{\mathbf{k}}\;.
\label{eq:M_SU2}
\end{eqnarray}
matrices (we keep the $m$ index only for bookkeeping purposes).
The  $m$ subspaces are spanned by the basis 
\begin{equation}
(\mathbf{t}_{m,{\bf k}}^{\dagger}, \mathbf{t}_{-m,-{\bf k}}^{\phantom{\dagger}}) ,
\end{equation}
defined in Eq.~(\ref{eq:tdmkdef}). 
Each of these operators is going to change the total $S^z_T$ by $m$, either by creating an $m$ triplon with $\mathbf{t}_{m,{\bf k}}^{\dagger}$ or by annihilating a $-m$ triplon with $\mathbf{t}_{-m,-{\bf k}}$. A rotation $e^{-i \varphi S^z_T}$ by an angle $\varphi$ about the $z$ axis in the spin space manifests in a phase factor
\begin{equation}
(\mathbf{t}_{m,{\bf k}}^{\dagger}, \mathbf{t}_{-m,-{\bf k}}^{\phantom{\dagger}}) \to e^{-i m \varphi} (\mathbf{t}_{m,{\bf k}}^{\dagger}, \mathbf{t}_{-m,-{\bf k}}^{\phantom{\dagger}}) .
  \label{eq:Szphase}
\end{equation}
Thereby in the Hamiltonian we encounter normal terms of the form 
$\mathbf{t}_{m,{\bf k}}^{\dagger} 
M_{m,\mathbf{k}}^{\phantom{\dagger}} 
\mathbf{t}_{m,{\bf k}}^{\phantom{\dagger}}$ 
and 
$\mathbf{t}_{-m,-{\bf k}}^{\phantom{\dagger}} 
M_{m,\mathbf{k}}^{\phantom{\dagger}}
\mathbf{t}_{-m,-{\bf k}}^{\dagger}$, 
and anomalous terms in the form of 
$\mathbf{t}_{m,{\bf k}}^{\dagger} 
N_{m,\mathbf{k}}^{\phantom{\dagger}}
\mathbf{t}_{-m,-{\bf k}}^{\dagger}$ 
and
$\mathbf{t}_{-m,{\bf k}}^{\phantom{\dagger}} 
N_{m,\mathbf{k}}^{\phantom{\dagger}}
\mathbf{t}_{m,-{\bf k}}^{\phantom{\dagger}}$ which are invariant with respect to the phase transformation of Eq.~(\ref{eq:Szphase}). These all commute with the 
\begin{equation}
S^z_T= 
 \mathbf{t}_{1,{\bf k}}^{\dagger} 
\mathbf{t}_{1,{\bf k}}^{\phantom{\dagger}}
-
\mathbf{t}_{-1,-{\bf k}}^{\dagger}
\mathbf{t}_{-1,-{\bf k}}^{\phantom{\dagger}} ,
\end{equation}
 so that the total $S^z_T$ is conserved, generating a U(1) symmetry. The higher SU(2) symmetry is exemplified by the matrices being independent from $m$.

\begin{figure}[h]
	\begin{center}
		\includegraphics[width=0.9\columnwidth]{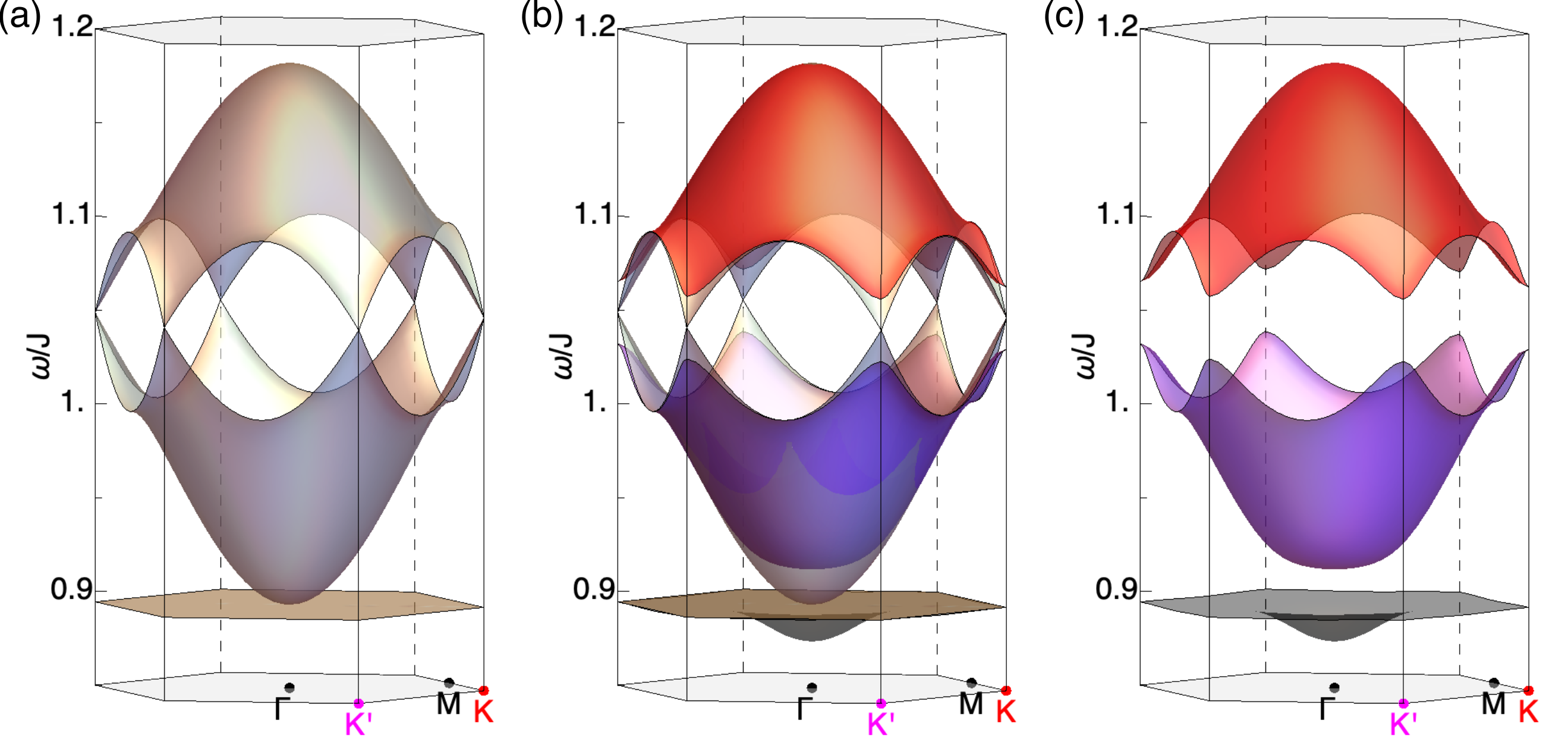}
		\caption{(a) Triplet bands in the SU(2) symmetric case, $J=1$, $J'=0.2$, $h=0$, and the DM and nematic terms are zero. All three bands are 3-fold degenerate due to the spin-rotation symmetry. (b) Triplet band structure for finite second neighbor DM interaction, $D''=0.01$.  (c) The $m=\pm1$ bands omitting the trivial $m=0$. All three bands are twofold degenerate, corresponding to the $+1$ and $-1$ spin of the triplets.}
		\label{fig:h0bands}
	\end{center}
\end{figure}

In this case, each of the three bands, coming from the three sublattices, are threefold degenerate, as the $m=1,0,-1$ have the same energies in the entire Brillouin zone. The dispersion computed from the SU(2) symmetric Bogoliubov--de Gennes Hamiltonian has the form 
\begin{subequations}
\label{eq:wmh0}
\begin{align}
   \omega_{1,m} &= \sqrt{J(J-2J')} \label{eq:w1mh0} \\
   \omega_{2,m} &= \sqrt{J(J+J')-\tilde J({\mathbf{k}})} \label{eq:w2mh0}\\
   \omega_{3,m} &= \sqrt{J(J+J')+\tilde J({\mathbf{k}})} \label{eq:w3mh0}\;.
 \end{align}
 \end{subequations}  
 with 
 $\tilde J({\mathbf{k}})=J  J' \sqrt{3+2\sum_\alpha\cos{\boldsymbol{\delta}_\alpha}\cdot\mathbf{k}}$.
We show these bands in Fig.~\ref{fig:h0bands}(a).
%
Additionally, discrete lattice symmetries give rise to degeneracies between the sublattice bands in the form of a linear band touching at the $K$ and $K'$ points, and a quadratic band touching at the zone center,~$\Gamma$.  

%
The intra-dimer Heisenberg coupling term, $J$, is nothing but the singlet--triplet gap, separating the triplet bands from the ground state, while the inter-dimer Heisenberg interaction, $J'$ gives dispersion to the triplets. 
The form of dispersion is immediately familiar from studies of graphene 
\cite{Haldane1988,Kane2005}, with the additional feature of a flat band 
just above the singlet--triplet gap, reflecting the frustration of the Kagome lattice.

From the band dispersion, Eq.~(\ref{eq:wmh0}), we can also read off the instabilities of the quantum paramagnet. 
For the ferromagnetic $J'\!=\!-J/4$, the $\omega_{3,m}$ softens at the $\Gamma$ point, signaling a transition to a long-range ordered time-reversal symmetry breaking state, where the spins on each layer are aligned ferromagneticaly, while  the two layers are aligned antiferromagneticaly. 
Similarly, we may notice that the energy of the $\omega_{1,m}$ flat band becomes 0 for antiferromagnetic $J'=J/2$, indicating a transition to a time-reversal symmetry breaking state, where the ordering is selected by the DM interactions or quantum fluctuations.

\section{DM interaction induced band topology }
\label{sec:band_topo}

In this Section, we explore the topologically non--trivial triplon bands 
which arise as result of DM interactions in a quantum paramagnet 
described by Eq.~(\ref{eq:H}), setting both the nematic terms and the 
magnetic field to zero. 
For simplicity, we will also set $J_\parallel = J_\perp = J$ in XXZ terms,  
such that they reduce to isotropic Heisenberg interactions.
The analysis is further simplified by the fact that
only the $z$ component of the DM interaction survives (up to first order).
It follows that total $S^z_T$  remains a good quantum number, and excitations with 
different $m=1,0,-1$ decouple from one another.


The out--of--plane DM interaction lowers the SU(2) symmetry to U(1), and splits the degeneracy at the corners and center of the BZ for the $m=\pm 1$ triplets, as shown in Fig.~\ref{fig:h0bands}(c). 
While $J'$ provides a real hopping amplitude, the inter-dimer DM interactions, $D'$ and $D''$ couple to the complex Gell-Mann matrices, and are responsible for the non-trivial topology, generating finite Berry curvature via the complex triplet hopping amplitude. 
The $m$-independent Heisenberg Hamiltonian~(\ref{eq:M_SU2}) is extended with the DM interaction~(\ref{eq:DM_hopp}). As the DM is diagonal with respect to the spin degrees of freedom, we can write the U(1)-symmetric Hamiltonian in a block-diagonal form, with decoupled $m=1,0,-1$ subspaces. We account for the $s^z$ operator in the DM interaction~(\ref{eq:DM_hopp}) with the factor $m$:
\begin{align}
M^{\text{U(1)}}_{m,\mathbf{k}} &= 
M^{\text{SU(2)}}_{m,\mathbf{k}}+
m \left[\!D' \! \cos\frac{\boldsymbol{\delta}_1\mathbf{k}}{2}\!+\!D'' \cos\!\frac{(\boldsymbol{\delta}_2\!-\!\boldsymbol{\delta}_3)\mathbf{k}}{2}\!\right]\!\lambda_5\nonumber\\
&\phantom{=} + m\left[\!D' \! \cos\frac{\boldsymbol{\delta}_2\mathbf{k}}{2}\!+\!D'' \cos\!\frac{(\boldsymbol{\delta}_3\!-\!\boldsymbol{\delta}_1)\mathbf{k}}{2}\!\right]\!\lambda_2\nonumber\\
&\phantom{=} + m\left[\!D' \! \cos\frac{\boldsymbol{\delta}_3\mathbf{k}}{2}\!+\!D'' \cos\!\frac{(\boldsymbol{\delta}_1\!-\!\boldsymbol{\delta}_2)\mathbf{k}}{2}\!\right]\! \lambda_7\;,\nonumber\\
N^{\text{U(1)}}_{m,\mathbf{k}} & = J I_3-M^{\text{U(1)}}_{m,\mathbf{k}}\;.
\label{eq:U(1)}
\end{align}
To determine the DM gap, we compute the energies at the $\Gamma$, $K$, and $K'$ points from the full Bogoliubov--de Gennes equations. 

At the $\Gamma$ point the bands have the energies 
\begin{subequations}
\label{eq:omegaG}
\begin{align}
&\omega_{1,m}(\Gamma)=\sqrt{J(J-2J'-m\Delta_\Gamma)}\;,\\
&\omega_{2,m}(\Gamma)=\sqrt{J(J-2J'+m\Delta_\Gamma)}\;,\\ 
&\omega_{3,m}(\Gamma)=\sqrt{J(J+4J')}\;,
 \end{align}
 \end{subequations}
 where $\Delta_\Gamma=2\sqrt{3}(D'+D'')$.  A finite  $\Delta_\Gamma$  opens a gap $\sqrt{J(J-2J'+\Delta_\Gamma)}-\sqrt{J(J-2J'-\Delta_\Gamma)}$ between the bands 1 and 2 at the $\Gamma$ point. 
 For $ \Delta_\Gamma \ll J,J'$, the gap becomes $\approx \sqrt{J/(J-2J')}\Delta_\Gamma $, thus proportional to the DM interactions.
 
At the $K$ and $K'$ points the bands have energies
\begin{subequations}
\begin{align}
&\omega_{1,m}(K)=\sqrt{J(J-2J')}\;,\\
&\omega_{2,m}(K)=\sqrt{J(J+J'+m\Delta_K)}\;,\\ 
&\omega_{3,m}(K)=\sqrt{J(J+J'-m\Delta_K)}\;,
 \end{align}
 \end{subequations}
where $\Delta_K=\sqrt{3}J(D'-2D'')$. This results in a gap of $\sqrt{J(J+J'+\Delta_K)}-\sqrt{J(J+J'-\Delta_K)}$ at the $K$ (and $K'$) point between the top bands with nonzero $m$. Note that the $m=0$ triplet band is not affected by the DM interaction, retaining their degeneracies at $K$, $K'$, and $\Gamma$ points. Furthermore, the $m=\pm 1$ triplets experience an opposite effect, reflecting the action of the spin-orbit coupling on the up and down spins, lending them an opposite torque. 
\begin{figure}[h!]
	\begin{center}
		\includegraphics[width=0.8\columnwidth]{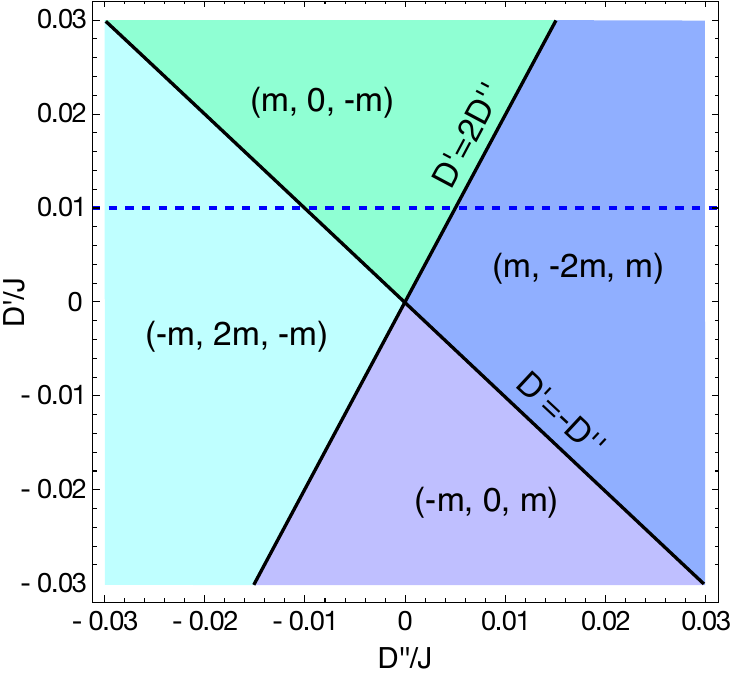}
		\caption{Chern numbers as the function of first ($D'$) and second ($D''$) neighbor inter-dimer DM interaction, listed in order of ascending band energy for a given $m$-subspace. The lines  $D'=2D''$ and $D'=-D''$ denote the boundaries of the band touching topological transitions. For $D'=2D''$ the gap between the upper two bands closes at the corners in the form of Dirac cones. Crossing the $D'=-D''$ line, the lower bands go through a quadratic touching at the $\Gamma$ point.}
		\label{fig:topo_PD}
	\end{center}
\end{figure}
\begin{figure*}[ht!]
	\begin{center}
		\includegraphics[width=6in]{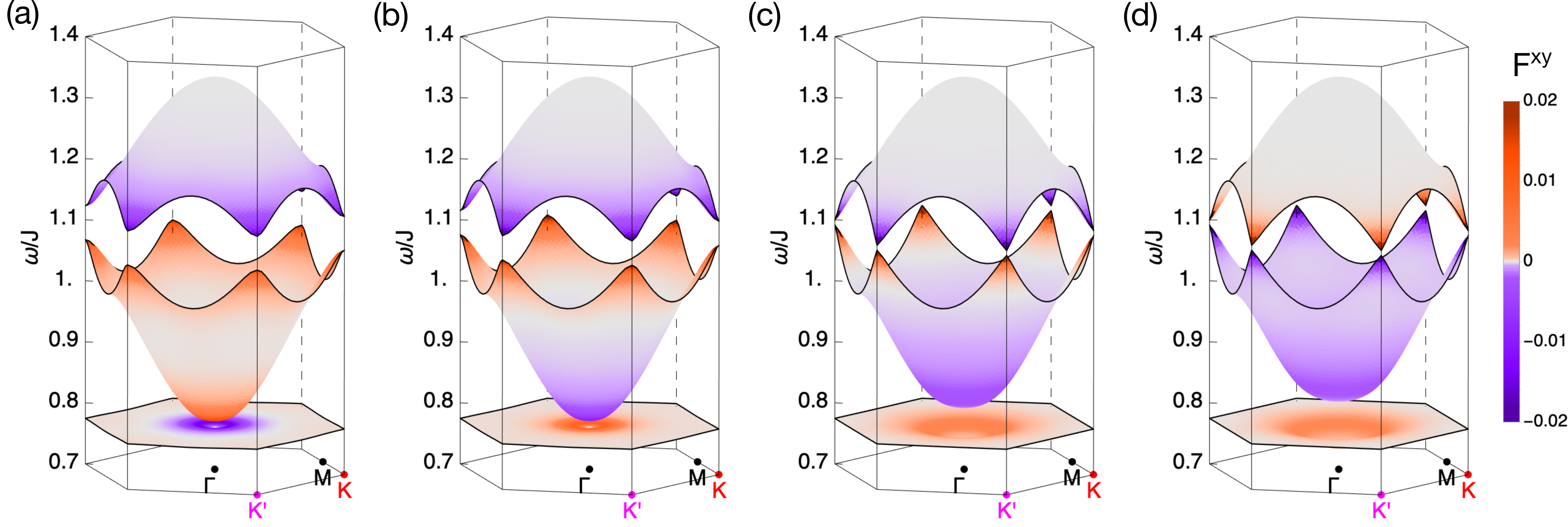}
		\caption{Distribution of Berry curvature in the vicinity of the transition points for the $m=1$ bands. The Berry curvature distribution for $m=-1$ is reversed, and not well definied for $m=0$. At $D'=-D''$ the gap closes at the $\Gamma$ point between the middle and lower bands. We chose the values $J=1$, $J'=0.2$, $D'=0.01$ (a) $D''=-0.01-\epsilon$ and (b) $D''=-0.01+\epsilon$, where $\epsilon=0.0025$. At $D'=2D''$ the gap at $K$ and $K'$ closes. Berry curvature for $D''=0.005-\epsilon$ (c) and $D''=0.005+\epsilon$ (d).}
		
		\label{fig:berrysCurvature}
	\end{center}
\end{figure*}
The three  triplet bands within the $m=1$ (or $m=-1$) subspace become fully gapped. Although the bands of $m=1$ are overlapping with the bands of $m=-1$ everywhere in the BZ, there are no matrix elements between the two, and (their canceling) Chern numbers can be computed independently for them~\cite{Simon1983,Berry1984,Kohmoto1985,Fukui2005}. The triplet bands in the case of a finite DM interaction are plotted in Fig.~\ref{fig:h0bands}(b) and (c). 
The Chern number of the $n$-th band of triplet $m$,
\begin{align}
C_{n,m}=\frac{1}{2\pi i}\int_{\rm BZ} dk_x dk_y F^{xy}_{n,m}\;,
\end{align}
 is the integral of the Berry curvature  
\begin{align}
F^{xy}_{n,m}({\bf k})\!\!=\!\!\left< \partial_{k_x} n({\bf k})| \Sigma^z \partial_{k_y} n({\bf k})\right>-\left< \partial_{k_y} n({\bf k})|\Sigma^z \partial_{k_x} n({\bf k})\right>
\label{eq:Berry_general}
\end{align}
over the Brillouin zone\cite{Shindou2013,Mook2014}. The $\left|n({\bf k})\right>$ eigenfunctions are the solutions of the Bogoliubov-de Gennes equations,
\begin{equation}
\begin{pmatrix}
  M^{\text{U(1)}}_{m,\mathbf{k}} & N^{\text{U(1)}}_{m,\mathbf{k}} \\
 N^{\text{U(1)}}_{m,\mathbf{k}} & M^{\text{U(1)}}_{m,\mathbf{k}} 
  \end{pmatrix}
\left|n({\bf k})\right>
 = \omega_{n,m}(\mathbf{k}) \Sigma^z \left|n({\bf k})\right> \;,
\end{equation}
 with
\begin{equation}
\Sigma^z = \begin{pmatrix}
I_3 & \mathbf{0} & \\
\mathbf{0} & -I_3 \\
\end{pmatrix} \;.
\label{eq:sigmazdef9x9}
\end{equation}

To map out a {\it band touching phase diagram}, we compute the Chern numbers as the function of the first and second neighbor DM interactions, $D'$ and $D''$ in Fig.~\ref{fig:topo_PD}. The numbers in Fig.~\ref{fig:topo_PD} represent the Chern numbers in the $m$ subspace, going from the bottom band to the top band. The $m=1$ and $m=-1$ triplet bands have opposite Chern numbers, reflecting their opposite chirality.

As the Chern number can only be changed via closing the gap, we can analytically determine the phase boundaries by examining when $\Delta_\Gamma$ or $\Delta_K$ become zero.

When $D'=2D''$, the $\Delta_K$ is zero closing the gap between the upper two bands at the $K$ and $K'$ points in a linear Dirac-cone-like touching. At each corner points, the Chern number of the bands is changed locally by one as the gap closes and reopens. The contribution from $K$ and $K'$ adds up to a $+2$ and $-2$ change in the Chern numbers of the upper and middle bands (see Fig.~\ref{fig:topo_PD}). The lowest band remains unaltered. 

$\Delta_\Gamma$ becomes zero for $D'=-D''$, closing the gap between the lower two bands in a quadratic touching at the zone center. The Chern numbers of the lower and middle bands change by $\pm 2$. The topmost band is unaffected. 

To illustrate the exchange of topological charge across the band touching transitions, we plotted the Berry curvature distribution on the $m=1$ triplet bands in the vicinity of  $D'=-D''$ and $D'=2D''$ in Fig.~\ref{fig:berrysCurvature}. The Berry curvature has opposite sign for the $m=-1$ triplet bands. 

When computing the Chern numbers shown in Fig.~\ref{fig:topo_PD}, and the distribution of the Berry curvature in Fig.~\ref{fig:berrysCurvature}, we use the numerical method introduced in Ref.~\cite{Fukui2005} equipped with the structure for particle-particle terms present in the Bogoliubov--de Gennes type of Hamiltonians. 
The numerical computation, however, does not provide us a deeper insight. Therefore, in Sec.~\ref{sec:expansion}, we give a detailed discussion on the topological transitions and the changing of the Chern numbers at the linear and quadratic touching points, restricting our analysis to the hopping part of the Bogoliubov--de Gennes Hamiltonian. The topological properties of the reduced hopping Hamiltonian and the Bogoliubov--de Gennes Hamiltonian are the same, as the $M$ and $N$ matrices are similar, i.e. they only differ in their diagonal.

\section{Linear versus quadratic band touchings}
\label{sec:expansion}

The transitions between bands with different Chern indices, shown in 
the phase diagram Fig.~\ref{fig:topo_PD}, occur where the gap between 
bands closes.
The point where bands touch, may have a linear or quadratic form, which we characterize in detail in this section. We provide analytic expressions for both 
the Berry curvature, and the associated Chern numbers, before and after the transitions, where they are well defined~\cite{Oshikawa1994}.
We focus on how topological charge is exchanged 
between bands.  
In both cases, two units of Chern number are exchanged.
For linear band touching this occurs through the exchange of a single unit 
at two different points in the BZ, $K$ and $K'$, while for quadratic band 
touchings two units are exchanged at the $\Gamma$ point.

We continue with the U(1) symmetric model of Eq.~(\ref{eq:U(1)}) including the isotropic Heisenberg interaction and the DM interactions, but keeping the nematic terms zero. 


Our approach will be to reduce the description of each touching 
point to a $2\times2$ matrix describing only those bands which touch
\begin{equation}
  M^{\rm eff}_{\mathbf{k}}= d_0(\mathbf{k}) I_2+\mathbf{d}(\mathbf{k})\cdot \boldsymbol{\sigma} \;,
\end{equation}
where $\boldsymbol{\sigma}$ is a vector of Pauli matrices, and the linear or quadratic 
form of the band touching is encoded in the 
coefficients $d_0(\mathbf{k})$ and $\mathbf{d}(\mathbf{k})$.
These coefficients can be expanded about the relevant wave vector, thereby 
allowing the analytic calculation of the Chern number 
\begin{equation}
 C = \frac{1}{2\pi} \int_{\text{BZ}} \Omega(k_x,k_y) dk_x dk_y\;
 \label{eq:Chern_def} 
\end{equation}
through the associated Berry curvature  
\begin{equation}
  \Omega(k_x,k_y) = \frac{1}{2} 
  \frac{\mathbf{d} \cdot \left(\frac{\partial \mathbf{d}}{\partial k_x} \times \frac{\partial \mathbf{d}}{\partial k_y}\right)}{(\mathbf{d}\cdot\mathbf{d})^{3/2}} ,
\label{eq:Berry}
\end{equation}
which, in this case, has the interpretation of a Skyrmion density.

For simplicity, the derivation we present is restricted to the hopping matrix $M^{\text{U(1)}}_{m,\mathbf{k}}$, and neglects the effect of the pair creation and annihilation terms.
This restriction can however be relaxed, at the expense of more lengthly expressions. If we were to consider the full Hamiltonian, the unitary transformations would correspond to $I_2\otimes U_\Lambda$, ($\Lambda=K,K',\Gamma$), and instead of the $2\times2$ matrix describing the bands in question, we would get a $4\times4$ problem that can be solved using the Bogoliubov transformation. 
The topological properties of the bands are generally not affected by the pairing terms (an exception is provided in Ref.~\cite{McClarty2018}, where the diagonal ($M$) and off-diagonal ($N$) matrices were not similar). The equivalency of the topology of $M$ and the full Hamiltonian has been 
discussed in Ref.~[\onlinecite{Nawa2018}].


Ultimately, our goal is to project the three-level problem (of each $m$ separately) onto a two-level one, involving the bands that touch at the corners and the center of the hexagonal Brillouin zone. To do this, we introduce unitary transformations, $U_{K}$ and $U_\Gamma$ that diagonalize $M^{\text{SU(2)}}_{\mathbf{k}}$ (and $N^{\text{SU(2)}}_{\mathbf{k}}$ too) at the band touching points. Naturally, we can perform the expansion about the touching point without projecting onto the two levels involved.  
Nonetheless, we restrict ourselves to a two-level problem to be able to represent the sublattice degrees of freedom with only three Pauli matrices instead of eight Gell-Mann matrices, which would give an eight-dimensional $\mathbf{d}$ vector.

\subsection{Linear band touching at $K$}

We begin with discussing the linear touchings that occur at the corners $\mathbf{k}_K \!= \!(\frac{4\pi}{3},0)$ and $\mathbf{k}_{K'} \!=\! (-\frac{4\pi}{3},0)$ of the Brillouin zone for $D'=2D''$. The matrix $U_K$ that diagonalizes $M^{\text{SU(2)}}_{\mathbf{k}}$ at both $K$-points is 

\begin{equation}
U_{K} =
\left(
\begin{array}{ccc}
 - \frac{1}{\sqrt{3}}&  \frac{1}{\sqrt{2}} & \frac{1}{\sqrt{6}}  \\
-  \frac{1}{\sqrt{3}} &- \frac{1}{\sqrt{2}}  &  \frac{1}{\sqrt{6}} \\
  \frac{1}{\sqrt{3}} & 0 &  \frac{2}{\sqrt{6}}  \\
\end{array}
\right) \;.
\end{equation}
The column vectors in $U_{K}$ correspond to the eigenvectors at K and also form a basis for a one- and two-dimensional irreducible representation of the three-fold symmetry at these points in the absence of the DM interactions. The band that belongs to the symmetric representation is the bottom band, which is well separated from the upper two bands in the vicinity of the zone-corners. The double representation stretches the subspace that we keep in the linearization.
Using $U_K$, we transform $M^{\text{U(1)}}_{m,\mathbf{k}}$ into the form 
\begin{widetext}
\begin{equation}
U_K^\dagger\cdot M^{\text{U(1)}}_{m,{\bf k}} \cdot U_K = 
  \left(
\begin{array}{ccc}
 J-J'& -\frac{1}{4} \sqrt{\frac{3}{2}} J' k_y+ i m \frac{3}{4 \sqrt{2}}D' k_x & \frac{1}{4} \sqrt{\frac{3}{2}}
   J' k_x+ i m \frac{3 }{4 \sqrt{2}}D' k_y \\
 -\frac{1}{4} \sqrt{\frac{3}{2}}  J' k_y-i m \frac{3}{4 \sqrt{2}}  D' k_x & J+\frac{ J'}{2}+\frac{ \sqrt{3} }{4} J'
   k_x & -\frac{ \sqrt{3} }{4} J' k_y+i m \sqrt{3} \left(\frac{D' }{2}-D'' \right) \\
 \frac{1}{4} \sqrt{\frac{3}{2}}  J'k_x-i m \frac{3}{4 \sqrt{2}} D' k_y& -\frac{\sqrt{3}}{4}   J' k_y-i m
  \sqrt{3} \left(\frac{D'}{2}- D'' \right) & J+\frac {J'}{2} -\frac{ \sqrt{3}}{4} J' k_x\\
\end{array}
\right),
\label{eq:UKMK}
\end{equation}
\end{widetext} 
where $m$ is the spin-index of the triplets taking the values $-1,0,1$, and $k_x$ and $k_y$ are measured from the $K$ point, i.e. $K$ corresponds to $k_x=k_y=0$. At the corners of the Brillouin zone, this matrix is block-diagonal and has eigen-energies $J-J'$, and $J+ J'/2 \pm m \sqrt{3} ( D'' - D'/2)$.  As we go away from the $K$ points, small off-diagonal matrix elements appear that are linear in momentum. Projecting~(\ref{eq:UKMK}) on the relevant subspace, we can write the two-band matrix as
\begin{equation}
\mathcal{H}^{\text{lin}}_{K}
 = \left(J+ \frac{J'}{2}\right) I_2  + \mathbf{d}_{K}\cdot \boldsymbol{\sigma}\;.
\end{equation}
The vector $\mathbf{d}_{K}$ has the form
\begin{subequations}
\begin{align}
d^x_{K} &= -\frac{\sqrt{3} }{4} J' k_y\;,\\
d^y_{K} &= -m \frac{\sqrt{3}}{2} \left(D'-2 D''\right)\;,\\
d^z_{K} &= \frac{\sqrt{3} }{4} J' k_x\;.
\end{align}
\label{eq:dK_vec}
\end{subequations}
The $\mathbf{d}_{K'}$ vector for the $K'$-point is given by $(d^x_{K'},d^y_{K'},d^z_{K'}) = (-d^x_{K},d^y_{K},-d^z_{K})$. 

\begin{figure}[b]
	\begin{center}
		\includegraphics[width=0.92\columnwidth]{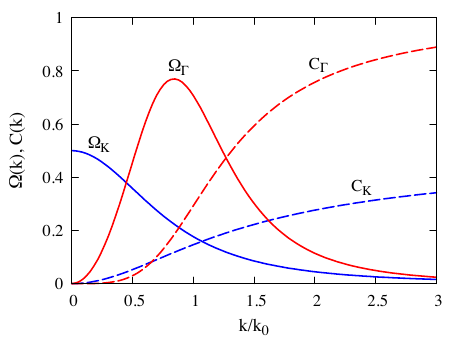}
		\caption{The radial distribution of the Berry curvature $\Omega(k)$ (solid lines) and the contribution to the Chern number $C(k)$ (dashed lines) for the linear band touching around the K points [Eqs.~(\ref{eq:BerryK}) and (\ref{eq:CkK})] and for the quadratic band touching around the $\Gamma$ point [Eqs.~(\ref{eq:BerryG}) and (\ref{eq:CkG})] in the Brillouin zone. The Berry curvature is maximal at the $K$ point for linear band touching, while in the case of the quadratic band touching it forms a ring--like structure around the $\Gamma$ point. In both cases, their integrals over a disk of radius $k$ centered at the touching points saturates quickly at the $1/2$ and $1$ values ($C(k)$).}
		\label{fig:BerryCurvatureGK}
	\end{center}
\end{figure}

Using Eqs.~(\ref{eq:Berry}) we get
\begin{equation}
  \Omega_K(k) =\frac{m J'^2(D'-2D'')}{( J'^2k^2+4(D'-2D'')^2)^{3/2}}\;,
\end{equation}
where we substituted $k^2_x+k^2_y=k^2$. We note that the Berry curvature does not depend on the valley index, {\it i.e.} it is the same at the $K$ and $K'$ points. To obtain a simpler form for $  \Omega(k) $, we introduce the dimensionless parameter 
\begin{align}
k_0=\frac{2(D'-2D'')}{J'}\;,
\end{align}
so that
\begin{align}
  \Omega_K(k) =m\frac{k_0}{2(k^2+k_0^2)^{3/2}}\;.
  \label{eq:BerryK}
\end{align}
$\Omega(k)$ has maximum at $k=0$, i.e. the Berry curvature is concentrated at the $K$ and $K'$ points, as shown in Figs.~\ref{fig:berrysCurvature} and~\ref{fig:BerryCurvatureGK}. 
Then  $\Omega_K =m\frac{{\rm sgn}(k_0)}{2 k_0^2}\propto  m \operatorname{sgn} (D'-2D'') \frac{ J'^2}{(D'-2D'')^2}$, which diverges as the $D'\to 2D''$. The $D'=2D''$ line marks the band touching transition at the two $K$-points as shown in Fig.~\ref{fig:topo_PD}. At this point the $d^y_{K}=0$ and the $\mathbf{d}_K$ vector is restricted to the $x$-$z$ plane. When we go around the $K$ point in the $(k_x,k_y)$ plane, the vector $(d^x_K,d^z_K)$ winds once around the origin.

Using Eq.~(\ref{eq:Chern_def}),
we integrate the Berry curvature Eq.~(\ref{eq:BerryK}) in a disk around the $K$ and $K'$ points that has radius $k$
\begin{align} 
 C_{K}(k) &= \frac{1}{2\pi} \int_0^{k}  \Omega_K(k') 2\pi k' dk' =
 \nonumber\\
 & = \frac m 2 {\rm sgn}(k_0) \left( 1 - \frac{ |k_0|}{\sqrt{k_0^2 + k^2}}\right) \;.
 \label{eq:CkK}
\end{align} 
In the vicinity of the transition point the second term goes to zero, and the $K$ and $K'$ points both contribute $\frac m 2$ to the total Chern number of the bands. As we cross the transition line $D'=2D''$, i.e. as the $k_0$ changes sign, the $\frac m 2$ Berry charge is exchanged and the Chern number is changed by $\pm 1$ both at K and K' -- the transferred charge is determined by the winding number at the touching point. The total change of the Chern number through the linear touching at the zone-corners is the sum of the contribution of $K$ and $K'$, corresponding to $\pm 2$, as indicated in Fig.~\ref{fig:topo_PD}. 

Let us note that $C_{K}$ does not give the total Chern number of the band calculated numerically in Fig.~\ref{fig:topo_PD}. $C_{K}$ only accounts for the Berry curvature concentrated around the $K$ (and $K'$) points, and does not include the contribution from the vicinity of the $\Gamma$ point, which is significant for the middle band as shown in Fig.~\ref{fig:berrysCurvature}. We use $C_{K}$ to discuss the exchange of topological charges through the band-touching transition at the corners of the BZ. For obtaining the total Chern number, one needs to consider the contribution of the Berry curvature at the zone corner too, which we discuss next.

\subsection{Quadratic band touching at $\Gamma$}

The quadratic touching is a little different. In the following, we will show that the total Berry charge is exchanged at a single point, where the Chern number changes by $\pm 2$. Thus, at the quadratic touching the bands have twice as much Berry charge as at the linear touching point. Furthermore, the Berry curvature in the case of the quadratic touching is not centered at a single point, as was the case with the linear touching. Instead, it is concentrated on a ring around the touching point. As the bands approach each other, the radius of the ring decreases, shrinking into a point when the bands touch. To see how this happens, we follow the procedure described above, expanding the rotated $M^{\text{U(1)}}_{m,{\bf k}}$ about the $\Gamma$ point.
The transformation matrix that diagonalizes the Heisenberg model $M^{\text{SU(2)}}_{m,{\bf k}}$ at the zone center has the form
\begin{equation}
U_{\Gamma} =
\left(
\begin{array}{ccc}
  \frac{1}{\sqrt{3}}&  \frac{1}{\sqrt{2}} & \frac{1}{\sqrt{6}}  \\
  \frac{1}{\sqrt{3}} &- \frac{1}{\sqrt{2}}  &  \frac{1}{\sqrt{6}} \\
  \frac{1}{\sqrt{3}} & 0 & - \frac{2}{\sqrt{6}}  \\
\end{array}
\right) \;.
\label{eq:UGamma}
\end{equation}
Rotating $M^{\text{U(1)}}_{m,{\bf k}}$ with~$U_{\Gamma}$, it becomes block-diagonal at the $\Gamma$ point, decoupling the touching bands from the top band. 
Moving away from the zone center, additional small matrix elements appear between the subspaces,
\begin{widetext}
\begin{equation}
U_\Gamma^\dagger\cdot M^m_{\Gamma} \cdot U_\Gamma \!=\!
\left(
\begin{array}{ccc}
 J-\frac{J'}{8}k^2+2 J' & -\frac{J'}{8 \sqrt{2}}k_x k_y & -\frac{J'}{16 \sqrt{2}}(k_x^2- k_y^2) \\
 -\frac{J'}{8 \sqrt{2}}k_x k_y & J-J'+\frac{J'}{8}k_x^2 & -i m \sqrt{3} (D'+ D'')+\frac{J'}{8} k_x k_y \\
- \frac{J' }{16 \sqrt{2}}(k_x^2-k_y^2) & i m \sqrt{3} (D'+ D'')+\frac{J'
  }{8} k_x k_y & J-J'+\frac{J' }{8}k_y^2 \\
\end{array}
\right) \;.
\label{eq:rotated_at_gamma}
\end{equation}
\end{widetext}
where $k^2_x+k^2_y=k^2$. We keep only the leading terms, taking $D',D'' \ll J'$ we neglect terms as $D'k^2$ and $D'' k^2$.
Diagonalizing this matrix at the $\Gamma$ point we find the energies $ \omega_{1} =J+2 J'$ for the upper band, and $ \omega_{2,3}= J-J' \pm m \sqrt{3}  ( D' + D'')$ for the lower bands split by the DM-interactions [cf. Eq.~(\ref{eq:omegaG})]. 
The effective $2\times 2$ Hamiltonian describing the splitting in the vicinity of the $\Gamma$ point is the bottom right corner of the matrix~(\ref{eq:rotated_at_gamma}),
\begin{equation}
  H_{\Gamma}^{\text{eff}} = J- J' \left( 1  - \frac{k^2 }{16} \right) I_2 +\mathbf{d}_{\Gamma} \cdot  \bm{\sigma} \;,
 \label{eq:HGamma}
\end{equation}
where the first term is an energy shift, and the $\mathbf{d}_{\Gamma}$ reads
\begin{subequations}
\begin{eqnarray}
  d^x_{\Gamma} &=& \frac{1}{16}    J'   2 k_x   k_y \;, \\
 d^y_{\Gamma} &=& m \sqrt{3} (D' + D'') \;,\\
 d^z_{\Gamma} &=& \frac{1}{16}    J'  ( k_x^2 - k_y^2 )  \;.
 \end{eqnarray}
 \label{eq:dvec}
 \end{subequations}
In the absence of the DM interactions  $d^y_{\Gamma}=0$, and we can recognize the mark of the quadratic band touching: as we go around the $\Gamma$ point in the $(k_x,k_y)$ plane, the vector $(d^x_\Gamma,d^z_\Gamma)$ winds twice around the origin \cite{Chong2008}. Turning on the $D'$ and/or the $D''$, a gap 
$\tilde \Delta_\Gamma(\mathbf{k}) = 2 \sqrt{\mathbf{d}_\Gamma\cdot\mathbf{d}_\Gamma}$ opens between the two bands of the size 
 \begin{equation}
 \tilde \Delta_\Gamma(\mathbf{k}) =  \sqrt{12 ( D' + D'' )^2 + \frac{1}{64}{J'}^2 k^4} ,
\end{equation}
[c.f. Eq.~(\ref{eq:omegaG})]. 
Inserting $\mathbf{d}_{\Gamma}(\mathbf{k})$ 
into Eq.~(\ref{eq:Berry}) for the Berry curvature of the lower band, we get 
\begin{equation}
   \Omega_\Gamma(k) = \frac{4 m \sqrt{3} (D' + D'') \left(\frac{J' k}{8}\right)^2}{\left[12 (D' + D'')^2 + \left(\frac{J' k^2}{8}\right)^2 \right]^{3/2}} \;,
\end{equation}
while $\Omega_\Gamma(k)$ changes sign for the upper band.
It is convenient to rewrite the curvature as
\begin{equation}
   \Omega_\Gamma(k) = m\frac{2  k^2    k_0^2 }{ \left( k_0^4  + k^4 \right)^{3/2}} \sgn (D'+D'')\;,
   \label{eq:BerryG}
\end{equation}
where 
\begin{equation}
  k_0^2 =  16\sqrt{3} \frac{|D' + D''|}{J'} \;.
\end{equation}
$\Omega_\Gamma(k)$ has a maximum for $k=2^{-1/4}k_0$, where it diverges  as 
\begin{equation}
 \Omega_\Gamma(2^{-1/4}k_0) \propto \frac{m}{k_0^2}\sgn (D'+D'') \propto m\frac{J'}{D'+D''}
\end{equation}
for $D' + D'' \to 0$.
Furthermore, the $\Omega_\Gamma(k)$ vanishes for both $k \ll k_0$ and $k \gg k_0$:
\begin{equation}
  \Omega_\Gamma(k) = 
  \begin{cases} 
  m \sgn (D'+D'') \frac{2 k^2}{k_0^4} + \cdots ,  
      &\mbox{if } k \ll k_0 ;\\ 
  m \sgn (D'+D'')  \frac{2 k_0^2}{ k^4} + \cdots, 
      & \mbox{if } k \gg k_0 .
  \end{cases}
\end{equation}
The maximum of the Berry curvature forms a ring around the $\Gamma$ point. The ring is nicely seen for the lowest two bands in Fig.~\ref{fig:berrysCurvature}. This behavior is unlike the linear band touching, where the Berry curvature is concentrated at the touching $K$-points.

Using Eq.~(\ref{eq:Chern_def}) and integrating the curvature around the $\Gamma$ point within a disk of radius $k$, we can check that the ring has enough curvature to collect a contribution $\pm 1$ to the Chern number of the bands:
\begin{align} 
 C_{\Gamma}(k)  & = m \sgn (D'+D'') \left(1 - \frac{ k_0^2}{\sqrt{k_0^4 + k^4}}\right) \label{eq:CkG}\\
  &\approx
    m \sgn (D'+D'') \left(1 - \frac{k_0^2}{k^2}\right) 
    , 
      & \mbox{if } k \gg k_0 .
\end{align}  
Here $m=-1,0,1$, and the sign depends on the band as well as the sign of the DM interaction. As the DM is continuously tuned across the the quadratic band touching transition, the $+1$ and $-1$ Berry charges are exchanged between the two bands, leading to the $\Delta C = 2$ transition for the two lowest bands, as seen along the $D' = -D''$ line in Fig.~\ref{fig:topo_PD} and~\ref{fig:berrysCurvature}. 

Again, we emphasize that investigating $ C_{\Gamma}$ characterizes how much the topological charge changes through band touching transition, and cannot produce the total Chern number of the band. For example, the middle band would have Berry curvature contribution from the corners of the BZ too, this however does not change when the gap closes at the $\Gamma$ point. 

%
\begin{figure}[b]
	\begin{center}
		\includegraphics[width=1\columnwidth]{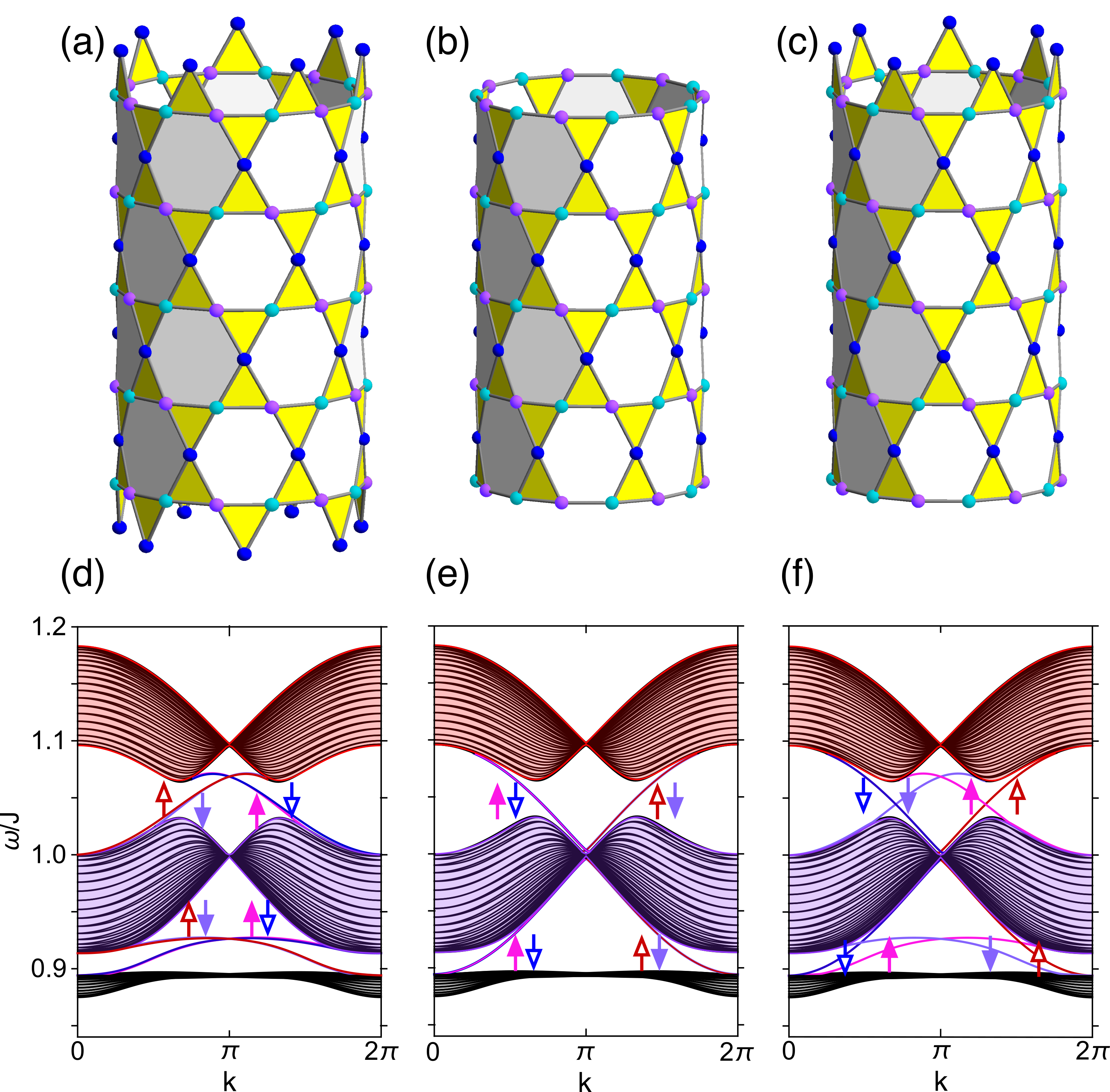}
		\caption{(a)--(c) Schematic figures representing periodic boundaries along the $x$ direction and a finite size in the $y$ direction with various edges. (d)--(f) The projected $m=\pm1$ triplet bands and helical triplet edge modes for finite second neighbor DM interaction, $D''=0.01$ in the open geometries shown in (a)--(c), respectively. The edge modes are colored according to the spin degrees of freedom and the edges, with red colors representing the up-spin and the blue colors denoting the down-spin. The open arrows correspond to the bottom edge, and the filled ones to the top edge.}
		\label{fig:open_geom}
	\end{center}
\end{figure}
\begin{figure*}[t!]
	\begin{center}
		\includegraphics[width=7in]{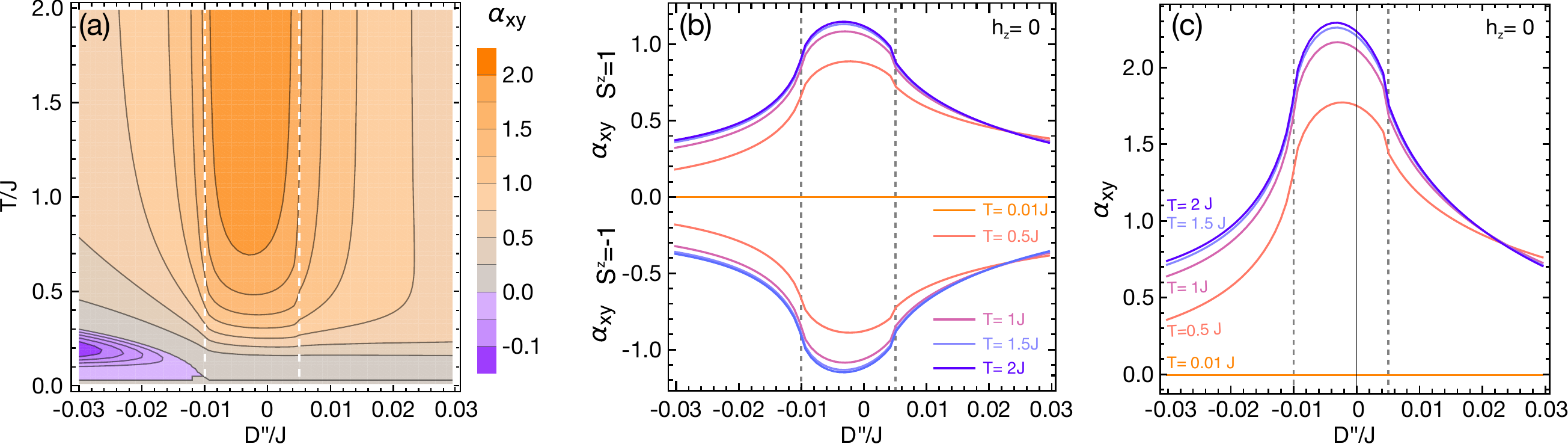}
		\caption{(a) The spin Nernst coefficient, $\alpha_{xy}$ along the blue dashed line of Fig.~\ref{fig:topo_PD} as a function of temperature and $D''$ ($J'/J=0.2$, $D'/J = 0.1$). The white dashed lines indicate the boundaries of the band touching topological transitions, where the Berry curvatures of the touching bands are exchanged. This change in the topological charge is also signaled by the anomaly of spin Nernst effect at the transition lines. 
(b) The contributions of the $S^z=+1$ and $-1$ triplet bands to the Nernst coefficient as the function of $D''$ at given temperatures. The $\alpha_{xy}$ has the same size and an opposite sign for the up and down spins, corresponding to their degenerate energies but opposite chirality. (c) The total $\alpha_{xy}=\alpha_{1,xy}-\alpha_{-1,xy}$ Nernst coefficient as the function of $D''$ at various temperature values. 
		}
		\label{fig:alpha_T_Dz}
	\end{center}
\end{figure*}

\section{Triplet $\mathcal{Z}_2$ topological insulator}
\label{sec:tripletZ2TopologicalInsulator}

We now turn to the problem of classifying the topological phases of the model,
and characterizing their experimental characteristics.
We start with the gapped, topological phase found in the simplified model 
of Section~\ref{sec:band_topo}, which we show to be a spin--Hall state 
characterized by a $\mathcal{Z}_2$ topological invarient, in direct analogue with
the model of Kane and Mele \cite{Kane2005}.

Within each $m$ subspace, we can think about the DM interaction term as an effective magnetic field in momentum space that acts on a pseudo spin-1 corresponding to the sublattice degree of freedom. The $m=0$ subspace is clearly no subject to any kind of magnetic field, while the spinful $m=1$ and $-1$ triples are affected in the opposite way, the effective field (DM) splits the "spin-1", both $m=1$ and $m=-1$ corresponding to the Haldane model~\cite{Haldane1988} with opposite chirality. Therefore, the time-reversal pair $m=1$ and $m=-1$ together realize an analog of the Kane and Mele model. Let us emphasize that the triplets, being components of an integer spin, do not form Kramer's pairs, and their degeneracy may be lifted even when the TR symmetry is preserved. 
When the nematic exchange anisotropies are present, for example, the $S^z_T$ ceases to be a good quantum number, and the $m=1$ and $m=-1$ subspaces hybridize, destroying the $\mathcal{Z}_2$ phase. We discuss this scenario in Sec.~\ref{sec:bad_nematic!}

In the case when the nematic terms are zero, the complete overlap between the $m=1$ and $m=-1$ bands renders the net Chern number zero. The $m=\pm1$ triplets realize an analog of the spin Hall insulator state and are characterized by a $\mathcal{Z}_2$ invariant~\cite{Kane2005,Kane2005b,Hasan2010,Qi2011}.

As shown in Sec.~\ref{sec:band_topo}, the Chern numbers are multiples of $m$ (see Fig:~\ref{fig:topo_PD}), and the bands with opposite spin have opposite Chern number. Therefore, the total Chern number of each degenerate band, formed by the time-reversal partners, $m=\pm 1$, vanishes: $C_{n,1} + C_{n,-1}=0$. Similar to electronic systems with conserved $S^z_T$, the $\mathcal{Z}_2$ index can be understood as the ``spin Chern number''~\cite{Kondo2019}  and computed as the staggered quantity $\frac{1}{2}(C_{n,1} - C_{n,-1})\mod 2$.

For the bottom and top  bands the Chern numbers in each phase are $\pm m$, thus the spin Chern number is  $\pm \frac{2}{2} \mod 2=1$. The middle band has either Chern number $0$, or $\pm 2m$, resulting in a trivial $0$ $\mathcal{Z}_2$ index. This also shows that the $\mathcal{Z}_2$ topological invariant does not depend on the DM anisotropy, but only on the conservation of $S^z_T$. Even when we close the band gaps at the band touching transitions, the $\mathcal{Z}_2$ indices will not change. The $\mathcal{Z}_2$ index can be computed using the eigenvalues of the parity operator too, which we discuss in detail in the Appendix~\ref{app:z2_from_parity}.

As a consequence of the $\mathcal{Z}_2$ topology, the system with open boundaries has helical triplet edge-modes, as shown in Fig.~\ref{fig:open_geom}. Note that one helical edge state is made of two chiral edge states going in opposite directions. We chose three different edge geometries, illustrated in  Fig.~\ref{fig:open_geom} (a)--(c), and computed the bands for each of those (see Fig.~\ref{fig:open_geom} (d)--(f)). The spin degree of freedom of the edge-modes is denoted with red and blue colors, while the filled and open arrows corresponds to the top and bottom edges, respectively.

\subsection{Triplet Nernst effect}

%
To obtain an experimentally detectable signature of the $\mathcal{Z}_2$ triplet bands, we compute the boson analog of the spin Hall effect. Applying a temperature gradient on the sample induces an energy current of triplet excitations. The $m=1$ and $-1$ triplets are affected in an opposite way by the DM interaction, due to their opposite chirality, and deflect into opposite directions. The triplet spin separation perpendicular to the temperature gradient leads to the cancelation of the transverse triplet heat current, but gives a finite transverse spin current. The transverse spin current arising in response to an applied temperature gradient is called the Nernst effect. We directly apply the formula of magnon mediated spin Nernst effect~\cite{Cheng2016,Zyuzin2016,Kovalev2016,Nakata2017,Zyuzin2018} for the triplet excitations, 
\begin{eqnarray}
j_{\text{SN}} = \alpha_{xy} \mathbf{\hat z} \times \nabla T \; ,
\end{eqnarray}
where the spin Nernst coefficient, $\alpha_{xy}$ can be expressed as
\begin{eqnarray}
\alpha_{xy}=-i \frac{k_B}{\hbar}\sum_{m,n}\int_{\rm BZ} m\cdot c_1(\rho_{n,m})F^{xy}_{n,m}(\mathbf{k})d^2\mathbf{k}\;,
\end{eqnarray}
where
$F^{xy}_{n,m}(\mathbf{k})$ is the Berry curvature of the $n$-th band of triplet $m$ defined in Eq.~(\ref{eq:Berry_general})
\begin{equation}
 c_1(\rho)=\int_0^\rho  \ln(1+t^{-1}) dt = (1+\rho) \ln(1+\rho) - \rho \ln \rho \,
 \end{equation}
and 
\begin{eqnarray}
\quad\rho_{n,m} = (e^{\omega_{\!n\!,\!m\!}\beta}-1)^{-1}
\end{eqnarray}
is the Bose--Einstein distribution function. 
 
A density plot of the numerically computed triplet mediated spin Nernst coefficient, $\alpha_{xy}$ is shown in Fig.~\ref{fig:alpha_T_Dz} as function of temperature and $D''$. We calculated $\alpha_{xy}$ along the blue dashed line in Fig.~\ref{fig:topo_PD}, using the complete Bogoliubov--de Gennes Hamiltonian. At  the topological band touching lines $D''=-D'$ and $D''=2D'$, the spin Nernst effect has inflection points, corresponding to the exchange of topological charge between the touching bands. 

As long as the magnetic field $h_z$ is zero and the time reversal symmetry is preserved, $\alpha_{xy,1}=-\alpha_{xy,-1}$, and the transverse spin Nernst current, $j_{\text{SN}}=j_{\text{SN},1}-j_{\text{SN},-1}$, can be written as $2j_{\text{SN},1}$. An applied magnetic field Zeeman-splits the triplets, pushing the $m=1$ and $m=-1$ bands in opposite directions (see Sec.\ref{sec:tripletChernInsulator}). As a consequence, the thermal filling of $m=1$ and $m=-1$ becomes different, leading to an imbalanced contribution from the up and down spins but still providing a finite spin Nernst effect. Note that we consider an out-of-plane field direction, that does not harm the U(1) symmetry, preserving $S^z_T$ as good quantum number.

\section{Nematic interaction and the fate of the $\mathcal{Z}_2$ phase}
\label{sec:bad_nematic!}

We now explore the consequences of the terms which mix the triplets 
with $m = \pm 1$, namely the nematic interactions introduced in Section~\ref{sec:symm_Hamilton}.
Symmetric exchange anisotropies of this type are naturally present in many spin systems, 
and arise in both the single-- and bilayer kagome and honeycomb models,  
widely discussed as candidates for topological magnon and triplon phases.


In the original Kane and Mele model~\cite{Kane2005}, analagous spin--mixing 
terms, such as the Rashba spin-orbit coupling, can be present 
(when the $\sigma_h$ reflection is broken). 
Although such a term hybridizes the bands with up and down spins, 
the spin-degeneracy at the time reversal invariant momenta (TRIM) 
remains protected by Kramer's theorem.
As a consequence, the $\mathcal{Z}_2$ topological phase considered 
by Kane and Mele is {\it perturbatively} stable against the introduction of Rashba interactions.


To explicitly see the difference from the Kane-Mele model,
we consider the subspace of the pseudo-spin-half formed by the $+1$ 
and $-1$ triplets, and investigate the fate of the $\mathcal{Z}_2$ band-topology 
when the spin-mixing nematic interactions are finite. 
We first examine their effect on the spin-degeneracy at the TRIM [Section~\ref{sec:TRIM}].
Then, taking an open geometry, we examine the consequence 
of the nematic interactions for the non-trivial edge modes  [Section~\ref{sec:edge.modes}].
To formalize our findings, in Section~\ref{sec:U1}
, we show that the TR$\times$U(1) symmetry corresponds to a pseudo time-reversal operator $\Theta$, which squares to $-1$ and can protect the $\mathcal{Z}_2$ band-topology. We identify the terms in our model that possess this symmetry and those which break it.
%


\subsection{Gap opening at the TRIM}
\label{sec:TRIM}

We consider first the effect of nematic interactions on triplon dispersion at 
time--reversal invariant momenta (TRIM). 
The Bogoliubov--de Gennes Hamiltonian of the TR-pair $m=\pm1$  triplets corresponds to the Hamiltonian~(\ref{eq:triplet_BdG}):
\begin{eqnarray}
H_{\mathbf{k}}^{(1,-1)}=
\left(\!\!
\begin{array}{cc}
M_{\mathbf{k}}& N_{\mathbf{k}} \\
N_{\mathbf{k}} & M_{\mathbf{k}}
\end{array}\!\!\right)\;.
\label{eq:H_sigma}
\end{eqnarray}
We compute the energies of the bands at the TRIM, namely 
\begin{eqnarray}
\Gamma &=& (0,0) \; , \qquad 
\mathsf{M}_1 = \left(\pi, \frac{\pi}{\sqrt{3}}\right) \; , \nonumber\\  
\mathsf{M}_2 &=& \left(0, \frac{2\pi}{\sqrt{3}}\right) \; , \qquad 
\mathsf{M}_3 = \left(-\pi, \frac{\pi}{\sqrt{3}}\right) \, 
\end{eqnarray}
shown in the bottom panel of Fig.~\ref{fig:BZ_and_deltas}(a).
Including the nematic terms, the energies at the $\Gamma$ point become 
\begin{subequations}
\label{eq:omegaGnem}
\begin{align}
\omega_{1,2}(\Gamma)=&J\!+\!\frac{J'}{2}\!-\!\frac{\sqrt{3}}{2}(D'\!+\!D'')-\Lambda,\\
\omega_{3,4}(\Gamma)=&J\!-\!J'\!+\!\sqrt{3}(D'\!+\!D'')\pm\frac{1}{2}(K_\| \!-\! 4K'_\|),\\ 
\omega_{5,6}(\Gamma)=&J\!+\!\frac{J'}{2}\!-\!\frac{\sqrt{3}}{2}(D'\!+\!D'')\!+\!\Lambda,
 \end{align}
 \end{subequations}
where 
 \begin{eqnarray}
 \Lambda=\frac{1}{2}\sqrt{3(D'+D''+\sqrt{3}J')^2+(K_\|+2K'_\|)^2}
\end{eqnarray} 
At the zone center, the top and bottom bands remain degenerate, while the 
middle two bands split as a result of the nematic interaction.

The effect of the nematic term is more drastic at the less symmetric $\mathsf{M}$ points, where the energies of the six triplets are all non-degenerate
\begin{subequations}
\label{eq:omegaM1}
\begin{align}
\omega_{1,2}(\Gamma)=&J\mp K'_\|-\Lambda_{\pm}\;,\\
\omega_{3,4}(\Gamma)=&J\mp\frac{K_\|}{2}\;,\\ 
\omega_{5,6}(\Gamma)=&J\mp K'_\|+\Lambda_{\pm}\;,
 \end{align}
 \end{subequations}
 with 
 \begin{eqnarray}
 \Lambda_{\pm}=\frac{1}{2}\sqrt{(2J'-\frac{K_\|}{2})^2+(2(D'-D'')\pm \frac{\sqrt{3}}{2}K_\|)^2} 
 \; . 
 \end{eqnarray}
The opening of a gaps $\Delta \sim K$ at these TRIM implies that even 
 infinitesimal nematic interactions are effective in destroying the 
 $\mathcal{Z}_2$ band-topology.

We note that, for simplicity, the eigenvalues and band-gaps above have been 
calculated considering the $M_{\bf{k}}$ matrix and not the entire 
Bogoliubov--de Gennes Hamiltonian. 
Solving the Bogoliubov--de Gennes problem denies us a simple analytical form, however, the 
physics remains the same: the infinitesimal nematic terms break spin-degeneracy, 
opening a gap at the TRIM.
This contrasts with the Kane--Mele model, where Rashba 
coupling does not break the time--reversal symmetry protecting 
the degeneracy at the TRIM \cite{Kane2005b}.
And it follows that infinitesimal Rashba interaction cannot lift the 
degeneracy of Kramers doublets at TRIM.

We return to this point in Section~\ref{sec:U1}, where we analyze the 
symmetry protecting $\mathcal{Z}_2$ band topology in our model, 
and show that nematic interactions break this symmetry.
Before that, we examine another consequence of the loss of $\mathcal{Z}_2$ 
band topology, namely the hybridization of the edge states.

\subsection{Hybridization of the edge states}
\label{sec:edge.modes}

\begin{figure}[t!]
	\begin{center}
		\includegraphics[width=1\columnwidth]{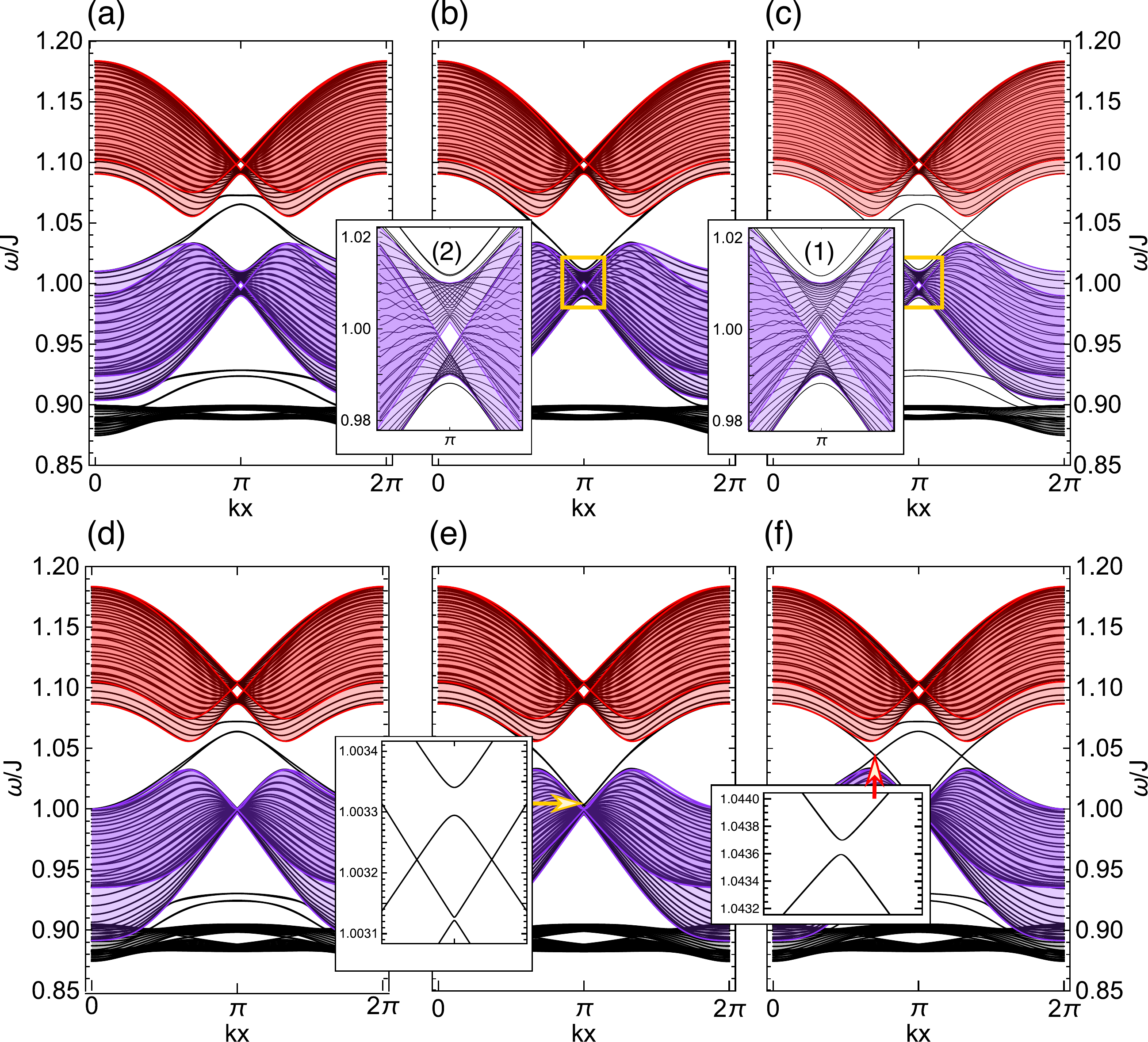}
		\caption{The effect of the nematic interaction on the edge modes and the topology of the bands. (a), (b), and (c) show the case when $J'=0.1 J$, $D''=0.01 J$, and $K_\|=0.02J$, for the spiky, flat, and mixed geometries, respectively (see Fig.~\ref{fig:open_geom} (a)--(c)). The $K'_\|$ is zero. The finite intra-dimer nematic term ($K$) leads to the hybridization of the helical edge modes with up and down spin, and a gap is opened at $k=\pi$ between the boundary-modes on the spiky edges, rendering them topologically trivial. The boundary modes at the flat edges do not merge with the middle band at $k=\pi$, but avoid that and become trivial as well as shown in the insets of (b) and (c).  (d), (e), and (f) represent the case when the inter-dimer nematic interaction is finite, $K'_\|=0.01J$ and $J'=0.1 J$, $D''=0.01 J$ but $K_\|=0$ . Here too the edge-modes with up and down spins hybridize and become gapped at $\pi$. In the case of the mixed boundaries, the edge-modes merge hybridization within the gap (see inset of (f)), and thus cannot collapse the band-gap, signaling topologically trivial bands.}
		\label{fig:nematic_open}
	\end{center}
\end{figure}

The band splitting at the TRIM already indicates that the degeneracy at these points are not protected by a symmetry corresponding to the time-reversal in the Kane-Mele model. 
To elaborate on the effect of the nematic terms on the $\mathcal{Z}_2$ topology, we compute the bands of the bilayer kagome stripes for the three different edge types shown in Fig.~\ref{fig:open_geom} (a)--(c). In an open system, the time reversal invariant point is $k=\pi$. The spin-degeneracy at this point will not necessarily be protected by TR symmetry, as it were for a Kramer's pair. To illustrate this, in Fig.~\ref{fig:nematic_open}, we plot the spinful triplet bands for the various  edge-geometries for finite intra-dimer ((a)--(c)) and finite inter-dimer nematic interaction ((d)--(f)). We find that in each case the edge modes hybridize, becoming trivial via an avoided crossing, i.e. they no longer connect the bands and close the gap. 
This is a clear indication that the fragile $\mathcal{Z}_2$ topology is quashed by the nematic terms. 

We note that the vanishing of  $\mathcal{Z}_2$ topology is not the consequence of a band-touching transition, which happens in the original Kane and Mele model, where increasing the Rashba term will close the band gap and induce a topological transition from a spin-Hall state to a trivial one. The values of the nematic terms were carefully chosen to be smaller than the critical values at which the band-gaps close (see Appendix~\ref{app:Kcestimates}).

\subsection{Pseudo-TR symmetry of the Hamiltonian}
\label{sec:U1}

In what follows, we give a more formal proof for the instability of the $\mathcal{Z}_2$ band-topology against an arbitrary small nematic interaction. For this, we will express the TR$\times$U(1) symmetry as a tensor product of $2\times 2$ and a $3\times 3$ matrix comprising the spin and sublattice degrees of freedom, respectively. 
The TR is an anti-unitary operator and in the basis $(\mathbf{t}_{{\bf k}}^{\dagger},\mathbf{t}_{-{\bf k}}^{\phantom{\dagger}})$ defined in Eqs.~(\ref{eq:basis}) has the form
\begin{equation}
T=(I_2 \otimes 2 s^x\otimes I_3)\cdot \mathcal{K}\;,
\label{eq:TRdef}
\end{equation}
where the 2-dimensional identity matrix $I_2$ accounts for the particle-hole space, $2 s^x=\sigma^x$ acts on the spin space spanned by $m=\pm 1$,  $I_3$ acts on the A, B, C sublattice degrees of freedom, and $\mathcal{K}$ is the complex conjugation. 
For details on the action of $T$ on the $+1$ and $-1$ triplets see Appendix~\ref{app:TR_symm}. We note that since the $+1$ and $-1$ triplets have integer spins the square of the TR is $T^2 = \mathbf{1}$, and it cannot ensure the Kramers degeneracy.

From our numerics it appears that the $U(1)$ symmetry --- the conservation of the $S^z_T$ --- is needed for the $\mathcal{Z}_2$ topology. In the $2 \times 2$ spin subspace the $U(1)$ rotation are described by the $e^{-i \varphi 2 s^z} \propto \cos(\varphi) I_2 - i \sin(\varphi) 2 s^z $, so the commutation of the bond wave Hamiltonian with the 
\begin{equation}
R= I_2 \otimes 2 s^z\otimes I_3 \;,
\label{eq:TRdef}
\end{equation}
unitary matrix ensures the $U(1)$ symmetry -- it commutes with the 
$M^{\text{XXZ}}_{\mathbf{k}}$, $M^{\text{DM}}_{\mathbf{k}}$, and $M^{\text{Zeeman}}_{\mathbf{k}}$  bond wave Hamiltonian matrices in Eq.~(\ref{eq:Msigmak}): $R\cdot\Sigma^z\cdot M^{\text{XXZ}}_{\mathbf{k}} - \Sigma^z\cdot M^{\text{XXZ}}_{\mathbf{k}}\cdot R = 0$, and so on, where we defined the pseudo-identity for the 
Bogoliubov--de Gennes formalism, 
\begin{equation}
\Sigma_z =2 s_z \otimes I_2 \otimes I_3 \;.
\end{equation}
The composition of the TR and U(1) symmetry gives the
\begin{equation}
  \Theta = (I_2 \otimes 2 i s^y \otimes I_3) \cdot \mathcal{K}  \\
\end{equation}
anti-unitary operator. Since
\begin{equation}
\Theta^2 = -1 \;,
\end{equation}
$\Theta$ is the desired pseudo-time reversal operator, assuming the role of the TR symmetry and ensuring   the protection of the $\mathcal{Z}_2$ topology.
We note that an analogous pseudo-TR operator was previously 
introduced in Ref.~\onlinecite{Kondo2019} for magnetically ordered spins.
However, this operator was associated with the combination of TR and mirror 
symmetries, which squares to +1 in the present case.

In what follows, we explicitly show that the Hamiltonian 
containing only the XXZ and DM anisotropies, $M^{\text{XXZ}}_{\mathbf{k}}$ and $M^{\text{DM}}_{\mathbf{k}}$ in Eq.~(\ref{eq:Msigmak}), commutes with $\Theta$ and will possess the $\mathcal{Z}_2$ topology. 
The nematic interactions, defined by Eq.~(\ref{eq:Mpm1_Nematic}), 
however, do not commute neither with $S^z_T$, nor with  $\Theta$. 
As a consequence, the $\mathcal{Z}_2$ bands will no longer be protected when either 
of the nematic interactions are finite. 

For this purpose we introduce the anti-unitary operator 
\begin{eqnarray}
	\mathcal{T}=(i 2 s^y\otimes I_3)\cdot \mathcal{K} \; , 
	\label{eq:defcalT}
\end{eqnarray}
which acts as a pseudo TR operator within the particle (hole) 
subspace:
\begin{eqnarray}
\Theta= I_2 \otimes\mathcal{T}=\left(\!\!
\begin{array}{cc}
 \mathcal{T}&\mathbf{0} \\
\mathbf{0} &  \mathcal{T}
\end{array}\!\!\right) .
\end{eqnarray}
The $\Theta$ antiunitary operator is a symmetry of $H_{\mathbf{k}}^{(1,-1)}$ if \cite{Kondo2019}
\begin{equation}
\Theta \Sigma^z H_{\mathbf{k}}^{(1,-1)}-  \Sigma^z H_{\mathbf{k}}^{(1,-1)}\Theta =0 \;.
\end{equation}
Using Eq.~(\ref{eq:defcalT}), this can be expressed as commutations between $\mathcal{T}$ and the diagonal $M_{\mathbf{k}}$ and  the off-diagonal $N_{\mathbf{k}}$ :
\begin{equation}
\Theta \Sigma^z H_{\mathbf{k}}^{(1,-1)}-  \Sigma^z H_{\mathbf{k}}^{(1,-1)}\Theta = \Sigma^z 
\begin{pmatrix}
\left[ \mathcal{T}, M_{\mathbf{k}}\right] &\left[ \mathcal{T}, N_{\mathbf{k}}\right]\\
\left[ \mathcal{T},N_{\mathbf{k}}\right]& \left[ \mathcal{T},M_{\mathbf{k}}\right]
\end{pmatrix}
\end{equation}
Since the matrices $M_{\mathbf{k}}$ and $N_{\mathbf{k}}$ only differ in their diagonal, it is sufficient to consider the commutation of $\mathcal{T}$ and $M_{\mathbf{k}}$. 
The various terms in  $M_{\mathbf{k}}$ are expressed in the same basis as $ \mathcal{T}$ in terms of $I_2\otimes I_3$, $s^\alpha \otimes I_3$, $I_2\otimes \lambda_n$, and $s^\alpha\otimes\lambda_n$, with $\alpha=x,y,z$, and $n=1,\hdots, 8$, as introduced in Sec.~\ref{sec:bond.wave.Hamiltonian}.
Let us note that the coefficients of these operators only contain even functions of ${\bf k}$, therefore $\mathcal{T}$ does not affect those.  

The effect of the $\mathcal{T}$ operator on a general term can be written as
\begin{subequations}
\begin{align}
\mathcal{T} (s^\alpha\otimes\lambda_n) \mathcal{T}^\dagger&=(i 2 s^y \otimes I_3)(s^\alpha\otimes\lambda_n)^*(i 2 s^y \otimes I_3)^\dagger\nonumber\\
&=(i 2 s^y) (s^\alpha)^* (i 2 s^y)^\dagger  \otimes (I_3)( \lambda_{n})^*(I_3)^\dagger\nonumber\\
&=(-s^\alpha)  \otimes \lambda_{n}^*,
\\
\mathcal{T} (I_2\otimes\lambda_n) \mathcal{T}^\dagger&=I_2  \otimes  \lambda_{n}^*,\\
\mathcal{T} (s^\alpha\otimes I_3) \mathcal{T}^\dagger&=-s^\alpha \otimes I_3,
\end{align}
\end{subequations}
where the $^*$ stands for complex conjugation. The $I_2\otimes I_3$ is trivially invariant. Furthermore, the operators $I_2\otimes \lambda_n$ preserve (break) the pseudo TR symmetry when the $\lambda_n$ Gell-Mann matrices are real (imaginary), while the terms $s^\alpha\otimes \lambda_n$ are invariant if $\lambda_n$ is imaginary and break $\mathcal{T}$ for $\lambda_n\in \mathbb{R}$. 

Table~\ref{tab:discrete_symm} contains the transformation of the different terms appearing in the triplet Hamiltonian under the pseudo and real TR operators and the U(1) symmetry. We can readily see that the nematic interactions break the U(1) symmetry and consequently the pseudo TR symmetry too that would guarantee the protection of the $\mathcal{Z}_2$ topology. The Zeeman term breaks both the pseudo and the physical TR symmetries. 

\begin{table}[htp]
\caption{Invariance of the various terms appearing in the Hamiltonian for the $m=\pm 1$ triplets under the pseudo time-reversal operator, the physical time-reversal symmetry, and the U(1) symmetry. Note that the pseudo TR symmetry corresponds to TR$\times$U(1).}
\label{tab:discrete_symm}
\begin{center}
\begin{ruledtabular}
\begin{tabular}{ccccc}
 \multirow{1}{*}{\rotatebox[origin=c]{90}{\parbox[c]{0.6 cm}{\centering }}} & &  Pseudo TR  & Physical TR & U(1) \\
 \multirow{1}{*}{\rotatebox[origin=c]{90}{\parbox[c]{0.6 cm}{\centering }}} & & $(i 2 s^y\!\otimes\! I_3) \mathcal{K}$ & $(2 s^x\!\otimes\! I_3) \mathcal{K}$ & $ 2 s^z\otimes I_3$  \\[0.7ex]
\hline
\multirow{3}{*}{\rotatebox[origin=c]{90}{\parbox[c]{1.4  cm}{\centering XXZ}}} & $I_2\otimes I_3 $  & $\checkmark$ & $\checkmark$ & $\checkmark$\\[1ex]
& $I_2\otimes\lambda_n$  & \multirow{2}{*}{$\checkmark$} & \multirow{2}{*}{$\checkmark$} & \multirow{2}{*}{$\checkmark$} \\
& (for real $\lambda_n$)  &  &  & \\[0.7ex]
\hline
\multirow{2}{*}{\rotatebox[origin=c]{90}{\parbox[c]{0.9  cm}{\centering DM}}}  & $s^z\otimes\lambda_n$   & \multirow{2}{*}{$\checkmark$}  & \multirow{2}{*}{$\checkmark$}  & \multirow{2}{*}{$\checkmark$} \\
& (for imaginary $\lambda_n$)  &  &  & \\[0.7ex]
\hline
\multirow{4}{*}{\rotatebox[origin=c]{90}{\parbox[c]{1.8  cm}{\centering Nematic}}} & $s^x\otimes\lambda_n$   &  \multirow{2}{*}{$-$}  &  \multirow{2}{*}{$\checkmark$}  & \multirow{2}{*}{$-$}\\
& (for real $\lambda_n$)  &  &  & \\[1ex]
 & $s^y\otimes\lambda_n$   &   \multirow{2}{*}{$-$}  &  \multirow{2}{*}{$\checkmark$}  & \multirow{2}{*}{$-$}\\
 & (for real $\lambda_n$)  &  &  & \\[0.7ex]
 \hline
  \multirow{3}{*}{\rotatebox[origin=c]{90}{\parbox[c]{1.1 cm}{\centering Zeeman}}} &   \multirow{3}{*}{$s^z\otimes I_3$} &   \multirow{3}{*}{$-$}  &   \multirow{3}{*}{$-$} &   \multirow{3}{*}{$\checkmark$}\\
  & & & & \\
  & & & &
\end{tabular}
\end{ruledtabular}
\end{center}
\label{default}
\end{table}%

\begin{figure}[h!]
	\begin{center}
		\includegraphics[width=0.9\columnwidth]{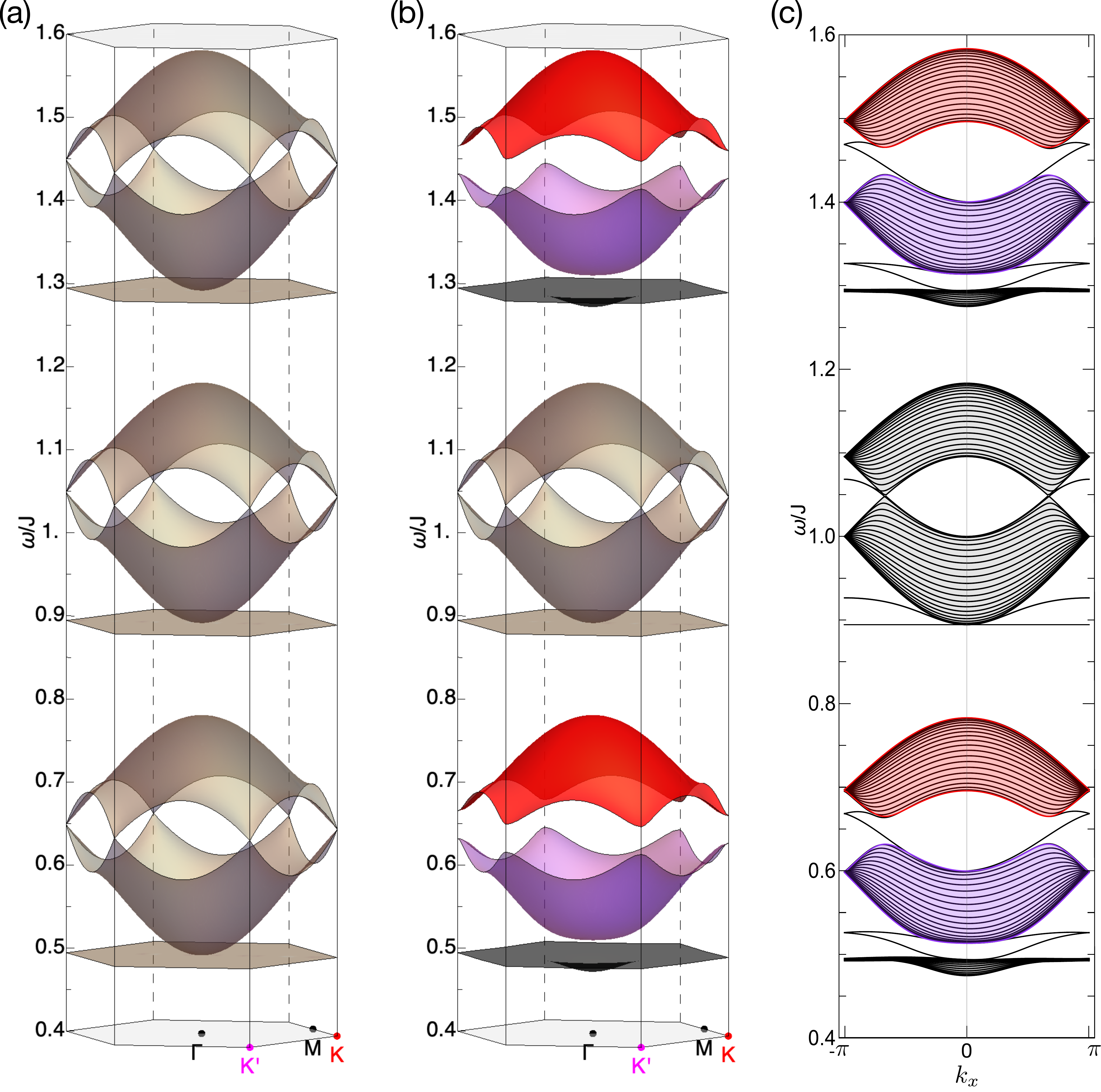}
		\caption{The finite magnetic field ($h = 0.2 J$, $g_z=2$) splits the triplets according to their spin degree of freedom $m$. (a) Triplet bands for the $D'=D''=0$ isotropic case. The bands of each $m$ sector have the same dispersion shifted by the Zeeman energy (b)$D'=0$, $D''=0.01 J$. The $m=\pm1$ bands  become fully gapped with well defined and finite Chern numbers, while the $m=0$ modes remain unaffected. (c) 1D dispersion in the open geometry (corresponding to Fig.~\ref{fig:open_geom}(a)) for the $D'=0$, $D''=0.01J$ case. Nontrivial edge states collapse the anisotropy gaps for $m=\pm1$, signaling nontrivial band topology. }
		\label{fig:spectrum_hz}
	\end{center}
\end{figure}

\begin{figure*}[ht!]
	\begin{center}
		\includegraphics[width=7in]{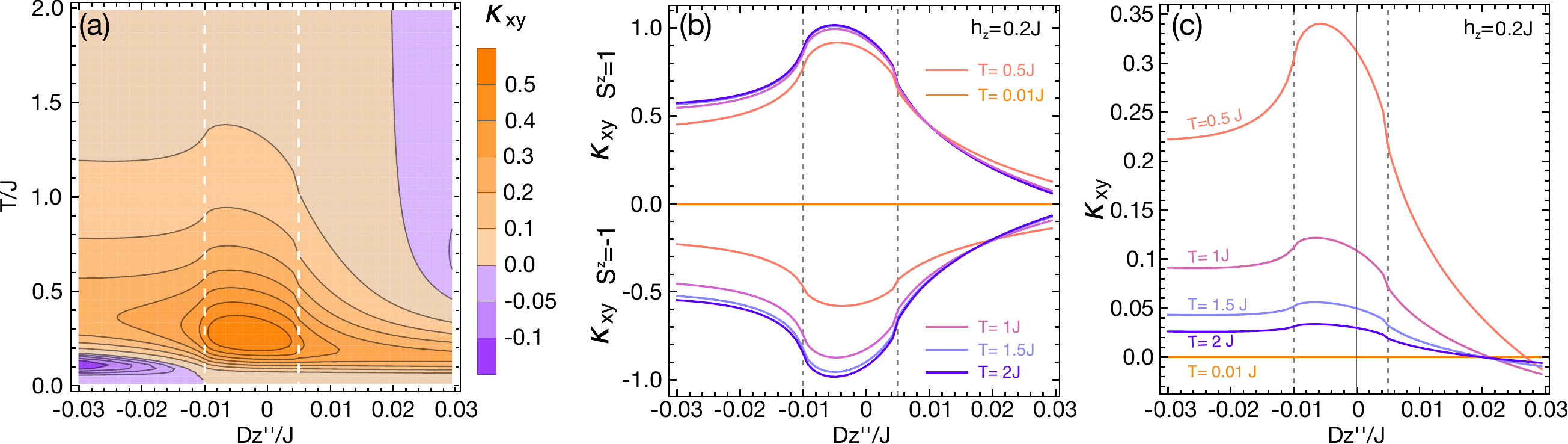}
		\caption{(a) The thermal Hall coefficient for $h = 0.2 J$ along the blue dashed line depicted in the phase diagram of \figref{topo_PD} using the same parameters as in Fig.~\ref{fig:alpha_T_Dz}. (b) We show the contributions from the $+1$ and $-1$ triplets separately. The $S^z=-1$ triplet moves up in energy, due to the Zeeman splitting, and thus at low temperatures these bands are less filled than those with $S^z=1$, giving a smaller contribution. At higher temperature this difference shrinks, and the sum of the signals, having opposite sign, cancels as $T\to\infty$. (c) Thermal Hall coefficient at various temperatures as the function of $D''$. The dashed vertical lines indicate the phase transitions at $D''=-D'$ and $D'' = D'/2$ respectively, where the thermal Hall coefficient has an inflection point. }
		\label{fig:thermalHallCoefficient}
	\end{center}
\end{figure*}
\section{Time-reversal symmetry breaking and thermal Hall effect}
\label{sec:tripletChernInsulator}

So far, we have only considered states found in the absence of magnetic field.
However the breaking of time--reversal symmetry by magnetic field also has interesting consequences.
In Fig.~\ref{fig:spectrum_hz}, we show how the triplon band structure of a model 
with Heisenberg and DM interactions changes as a function of magnetic field.

An immediate consequence of magnetic field is that the time--reversal pairs 
$m=\pm1$ split, and their Chern numbers no longer cancel.
As a result, we end up with an analog of a Chern insulator state, but 
realized by the triplets.

A second consequence of the Zeeman splitting of triplets is that the thermal filling of $m=1$ and $m=-1$ becomes different, and the imbalance of the up and down spin current produces a finite thermal Hall coefficient. 
The thermal Hall signal is the transverse energy current in response to an applied temperature gradient (and perpendicular magnetic field). 
In the TR symmetric case the up and down-spin triplets had overlapping energies and consequently identical thermal filling, providing the same number of excitations moving in opposite directions and giving a zero net thermal Hall response.  
As the degeneracy is lifted, the up and down-spin contributions become different giving a finite net transverse energy current. 
The thermal Hall coefficient can be written as~\cite{Matsumoto2011}.
\begin{eqnarray}
\kappa_{xy}= -i\frac{1}{\beta}\sum_{n,m}\int_{\rm BZ} c_2(\rho_{n,m})F^{xy}_{n,m}(\mathbf{k})d^2\mathbf{k}\;,
\label{eq:kappaxy}
\end{eqnarray}
where $F^{xy}_{n,m}(\mathbf{k})$ is the Berry's curvature, $c_2(\rho)=\int_0^\rho \ln^2(1+t^{-1})dt$, and $\rho_{n,m}$ is the Bose-Einstein distribution.

We computed the thermal Hall coefficient, $\kappa_{xy}$ along the blue dashed line indicated in Fig.~\ref{fig:topo_PD} as the function of temperature and using the entire Bogoliubov--de Gennes Hamiltonian. $\kappa_{xy}$ is plotted in Fig.~\ref{fig:thermalHallCoefficient}. 

In summary, the parameters $D'$ and $D''$ together with the Chern numbers define a topological phase diagram as seen on \figref{topo_PD}. The different phases can be clearly distinguished by the thermal Hall coefficient which is largest in the $(m,0,-m)$ phase, as can be seen on \figref{thermalHallCoefficient}. The effect is equally large in the $(-m,0,m)$ phase, but of opposite sign, as this phase can be reached simply by inverting the sign of $D'$ and $D''$ thereby negating the sign of the $m$-dependent term in the Hamiltonian.

Let us point out that including the nematic terms have no effect on the Chern bands, unlike the case of the $\mathcal{Z}_2$ bands. As long as the nematic interactions are small enough for the band-gap to remain open, the topology of the triplet bands does not change.
 
\section{Conclusions}
\label{sec:conclusions}

In this Article, we investigate  some of the novel features which arise in the topology of the triplon bands found in spin--1/2 quantum paramagnet on bilayer kagome lattice.
 We go beyond the XXZ model extended with the DM interactions, the archetypal analogue of the electronic tight-binding hopping Hamiltonian with spin-orbit coupling. Deriving the most general form of the Hamiltonian allowed by the symmetry 
of the lattice [Section~\ref{sec:symm_Hamilton}] we explore the ramifications of each symmetry-allowed terms. 
Reducing  the Hamiltonian
to a model for triplon excitations of the quantum paramagnet [Section~\ref{sec:modelHamiltonian}], we characterize these bands for models of increasing complexity, by lowering the symmetries.


The simplest case is when the Hamiltonian is the pure SU(2) symmetric Heisenberg model, discussed in Sec.~\ref{sec:spin.conserving.limit}. In this case, the band structure is trivial, with 3-fold degenerate bands, exhibiting a quadratic band touching at the $\Gamma$ point and linear band touchings at the $K$ and $K'$ points in the Brillouin zone.

In the TR$\times$U(1) symmetric case, we find that the 
triplets provide an analog to the $\mathcal{Z}_2$ topological insulator considered
by Kane and Mele~\cite{Kane2005,Kane2005b}, with helical triplet modes on open edges 
[Section~\ref{sec:band_topo}].
This model supports topological phases with bands having different Chern numbers. 
We give a detailed description of the exchanged topological charges at the phase transitions, for both the linear and quadratic touchings [Section~\ref{sec:expansion}].
The behaviour of the $\mathcal{Z}_2$ topological phase is 
also characterized through calculations of its topological invariant, 
and the associated triplet Nernst effect [Section~\ref{sec:tripletZ2TopologicalInsulator}].

Finally, we explore two different mechanisms which can eliminate 
the $\mathcal{Z}_2$ topology of the triplon bands by breaking either the pseudo-TR symmetry with the inclusion of nematic interactions, or by breaking the physical TR symmetry with applied magnetic field.
The first route to remove the $\mathcal{Z}_2$ phase is the breaking of time--reversal symmetry. In applied magnetic field the triplon bands split and the system becomes a Chern insulator, exhibiting finite thermal Hall response and chiral modes on open edges [Section~\ref{sec:tripletChernInsulator}], in a straightforward analogy with the Kane and Mele model \cite{Kane2005,Kane2005b}.
The second, less trivial route, is the inclusion of bond--nematic interactions, permitted by the 
symmetry of the lattice, which breaks a pseudo TR symmetry introduced in Section~\ref{sec:bad_nematic!}.

In contrast to the electronic model of Kane and Mele, where the mixing of states with 
$S^z = \pm \frac 1 2$ by (weak) Rashba interactions is compatible with a spin--Hall state,  
these terms have a singular effect, immediately changing the topology of 
the triplon bands. Such bond--nematic interactions, typically referred to as symmetric exchange anisotropies, are naturally present in other spin models too, proposed to exhibit $\mathcal{Z}_2$ bands realized by magnetic excitations. Although one can introduce a pseudo TR symmetry that squares to $-1$, in analogy to the physical TR symmetry present in the Kane and Mele model, this symmetry does not prevail in a general model.

The nematic terms and in-plane DM interactions that mix the spins and also break the pseudo TR symmetry ($\Theta$) can also be present in other models proposed as bosonic analogues of $\mathcal{Z}_2$ bands, including bilayer ordered magnets~\cite{Kondo2019}, and paramagnets~\cite{Joshi2019}. Our results call for a  detailed investigations of the consequences of the various $\Theta$ symmetry-breaking terms in bosonic systems in general.

\acknowledgements

%
The authors are pleased to acknowledge discussions with 
Christian D. Batista and Yutaka Akagi.
The authors also gratefully acknowledge the hospitality of the KITP program 
{\it ``Topological Quantum Matter: Concepts and Realizations''}, where 
a part of the work was carried out.
This work was supported by the Theory of Quantum Matter Unit of the 
Okinawa Institute of Science and Technology Graduate University (OIST), 
and the Hungarian NKFIH Grant No. K124176. 
This research was supported in part by the National Science Foundation under 
Grant No. NSF PHY-1748958.

\appendix

\section{Time-reversal symmetry}\label{app:TR_symm}

The TR operator for a dimer has the from 
\begin{eqnarray}
T=e^{i \pi (S^y_1+S^y_2)}\cdot \mathcal{K}=\left(\begin{array}{cccc}
1 & 0 & 0 & 0\\
0 & 0 & 0 & 1\\
0 & 0 & -1 & 0\\
0 & 1 & 0 & 0
\end{array}\right)\cdot  \mathcal{K}\;,
\end{eqnarray}
where the basis is $(\left| s\right>,\left| t_{-1}\right>,\left| t_{0}\right>,\left| t_{1}\right>)$ and $\mathcal{K}$ denotes a complex conjugation. 
Thus the TR operator for the spin-1 formed by the triplets is 
\begin{eqnarray}
T=\left(\begin{array}{ccc}
0 & 0 & 1\\
0 & -1 & 0\\
1 & 0 & 0
\end{array}\right)\cdot  \mathcal{K}\;.
\end{eqnarray}
The state $\left| t_{0}\right>$ changes sign, while the $\left| t_{1}\right>$ and $\left| t_{-1}\right>$ transform into each other. This can also be shown writing the triplets in their usual form   $\left| t_{1}\right>=\left| \uparrow\uparrow\right>$,  $\left| t_{-1}\right>=\left| \downarrow\downarrow\right>$,  and $\left| t_{0}\right>=\frac{1}{\sqrt{2}}(\left| \uparrow\downarrow\right>+\left|\downarrow\uparrow\right>)$,  and using that the TR acts on the spin-half as $T:\left| \uparrow\right>\to\left| \downarrow\right>$, and $T:\left| \downarrow\right>\to -\left| \uparrow\right>$. 

Restricting ourselves to the up and down triplet states, and considering them as the components of a pseudo spin-half, the Pauli matrices formed by them will transform differently under TR than those of a real spin-half (where all of them break TR). The TR acts on the pseudo-up and down spins as $T:\left| 1\right>\to\left| -1\right>$, and $T:\left| -1\right>\to \left| 1\right>$. Therefore, among the Pauli matrices 
\begin{subequations}
\begin{eqnarray}
\sigma^x&=& \left| 1\right>\left<-1\right| +\left| -1\right>\left<1\right| \\
\sigma^y &=&-i \left| 1\right>\left<-1\right| +i \left| -1\right>\left<1\right| \\ 
\sigma^z&=& \left| 1\right>\left<1\right| -\left| -1\right>\left<-1\right|\;,
\end{eqnarray}
\end{subequations}
the $\sigma^x$ and $\sigma^y$ are invariant under TR, and $\sigma^z$ remains the only TR breaking operator. This is an important difference between the pseudo-spin-half formed by the triplets $\left| 1\right>$ and $\left| -1\right>$, and the real spin-half of an electron, for example. 

We now write the $T$ operator in the basis of Eqs.~\ref{eq:basis}: First, we represent the action of $T$ on the $\left| 1\right>$ and $\left| -1\right>$ triplets with the $\sigma_x=2 s_x$ operator, then we account for the particle-hole space of the Bogoliubov--de Gennes equation by a 2-by-2 identity matrix $I_2$, and the three sublattice flavors by a 3-by-3 identity matrix $I_3$, since the TR operator leaves the sublattices intact. The TR operator becomes $T=(I_3\otimes 2 s^x \otimes I_3)\cdot \mathcal{K}$ where $\mathcal{K}$ is the complex conjugation. In the sublattice subspace represented by the Gell-Mann matrices, the real $\lambda$'s are TR invariant, while the imaginary $\lambda_2$, $\lambda_5$, and $\lambda_7$ break time reversal symmetry.

In the main text we discuss the U(1) symmetry operator written in the same basis as $T$, as well as the TR$\times$U(1) symmetry corresponding to a pseudo TR operator $\Theta$. We show that to see the commutation relation of the the various terms in the Bogoliubov--de Gennes Hamiltonian it is sufficient to consider the particle (hole) subspace. In Table~\ref{tab:discrete_symm}, we collect the transformation of the different terms entering the triplet Hamiltonian, $M_{{\bf k}}$. 

\section{$\mathcal{Z}_2$ invariant from the parity eigenvalues}\label{app:z2_from_parity}

In an inversion symmetric system, the $\mathcal{Z}_2$ can be  easily calculated using the parity eigenvalues at the four time-reversal invariant momenta (TRIM)~\cite{Fu2007}.
\begin{equation}
\prod_{i=1}^4\xi(\Gamma_i)=(-1)^\nu\;.
\label{eq:z2}
\end{equation}
The TRIM, $\Gamma_i$ correspond to $\Gamma$, $\mathsf{M_1}$, $\mathsf{M_2}$, and $\mathsf{M_3}$ shown in Fig.~\ref{fig:BZ_and_deltas}(a) of the main text. When the product is $-1$, the exponent $\nu$ is odd and the system is topologically nontrivial. For even  $\nu$ values the system is trivial.

We define the parity operator ($\mathcal{P}$) as the inversion through the center of the dimer $A$. Consequently,  the effect of $\mathcal{P}$ on dimer $A$ is the exchange of its sites $1$ and $2$, while dimer $B$ will also be shifted by ${\boldsymbol\delta}_y$ and dimer $C$ by ${\boldsymbol\delta}_x$ from their original positions beside exchanging their sites (see Fig.~\ref{fig:BZ_and_deltas} in the main text).

Changing the site indices, $1$ and $2$ does not affect the triplets, which are even under permutation, nor has the inversion any effect on the spin degrees of freedom. Thus, $\mathcal{P}$ will not mix different $m$ bands and we can treat each sector separately again.
In momentum space $\mathcal{P}_m$ becomes diagonal,
\begin{eqnarray}
\mathcal{P}_m\!\!=
\!\!\left(\!\!
\begin{array}{c}
\mathbf{t}_m^{\dagger}(\mathbf{k})\\
\mathbf{t}_{-\!m\!}^{\phantom{\dagger}}({-\!\bf k})
\end{array}\!
\right)\!\!
\left(\!\!
\begin{array}{cc}
P_\mathbf{k} & \boldsymbol{0}\\
\boldsymbol{0} & P_\mathbf{k}
\end{array}\!\!
\right)
\!\left(\!\!
\begin{array}{c}
\mathbf{t}_m^{\phantom{\dagger}}(\mathbf{k})\\
\mathbf{t}_{-m}^{\dagger}({-\!\bf k})
\end{array}\!
\right)\!\!\;\;
\label{eq:parity}
\end{eqnarray}
where 
\begin{equation}
P_\mathbf{k}={\rm diag}
\left(
1, e^{i {\boldsymbol\delta}_y\cdot \mathbf{k}}, e^{i {\boldsymbol\delta}_x\cdot \mathbf{k}}\right)
\label{eq:Ik}
\end{equation}

We compute the eigenvectors of $\mathcal{H}_m$ numerically at each TRIM, $\Gamma_i=(\Gamma, \mathsf{M_1}, \mathsf{M_2}, \mathsf{M_3})$, and determine their eigenvalue, $\xi_m(\Gamma_i)$ with $\mathcal{P}_m$ defined in Eq.~(\ref{eq:parity}). When inversion symmetry is present, the $\mathcal{P}_m$ commutes with $\mathcal{H}_m$ at the TRIM. The parity eigenvalues of the bands are collected in Table~\ref{tab:Z2}. The $\mathcal{Z}_2$ indices of the bottom and top bands are 1, while it is 0 for the middle band. Let us note that if we chose dimer $B$ as the center of inversion, the parity eigenvalues for the point $\mathsf{M_2}$ would become $-1$, $1$, and $-1$, while  $\xi_m(\mathsf{M_1})$ and  $\xi_m(\mathsf{M_3})$ would be $1$, $-1$, and $1$. Similarly, setting dimer $C$ as the inversion center results in a further cyclic permutation of the rows of Table~\ref{tab:Z2}. This corresponds to the three-fold symmetry of the $ABC$-triangles.
\begin{ruledtabular}
	\begin{table}[htp]
	\caption{Parity eigenvalues of the bottom (black), middle (purple) and, top (red) bands for each spin degree of freedom, $m$ at the time reversal invariant momenta, $\Gamma_i$.}
		\label{tab:Z2}

		\begin{center}
			\begin{tabular}{ccccc}
				band & & bottom & middle & top\\
				\hline
				\multirow{4}{*}{\rotatebox[origin=c]{90}{\parbox[c]{2 cm}{\centering parity eigenvalue}}} & $\xi_m(\Gamma)$ & 1 & 1 & 1\\[1ex]
				& $\xi_m(\mathsf{M_1})$ & -1 & 1 & -1\\[1ex]
				& $\xi_m(\mathsf{M_2})$ & 1 & -1 & 1\\[1ex]
				& $\xi_m(\mathsf{M_3})$ & 1 & -1 & 1\\
				\hline
				$\mathcal{Z}_2$ index  & $\nu$ & 1 & 0 & 1
			\end{tabular}
		\end{center}
			\end{table}%
\end{ruledtabular}

\section{Estimates of the gap closing transition induced by the nematic terms }
\label{app:Kcestimates}

%
In the $K'_\|=0$ case, corresponding to Fig.~\ref{fig:nematic_open} (a)--(c), the critical value of the intra-dimer nematic interaction at which the band-touching occurs is $K^{\rm crit}_{\|}=\frac{4 \left(\tilde{D}^2-\sqrt{3} \tilde{D} \tilde{J'}\right)}{\sqrt{3} \tilde{D}-\tilde{J'}}$. The parameters $\tilde{D}$ and $\tilde{J'}$ depend on where the gap closes in the Brillouin zone: The gap between two low-lying bands close at the $\Gamma$ point for $\tilde{D}\approx(D'+D'')(\frac{J'}{J}+1)$ and $ \tilde{J'}\approx J'-\frac{J'^2}{2 J}+\frac{(D'+D'')^2}{2 J}$. While, the gap between the top bands collapses at the $K(K')$ point for $\tilde{D}\approx(D'-2D'')(\frac{J'}{4 J}-\frac{1}{2})$ and $ \tilde{J'}\approx-\frac{J'}{2}-\frac{J'^2}{8 J}+\frac{(D'-2D'')^2}{8 J}$. The values of $\tilde{D}$ and $\tilde{J'}$ were determined perturbatively, assuming that $J$ is the leading term. Inserting the parameter values used in Fig.~\ref{fig:nematic_open} (a)--(c), we get $K^{\rm crit}_{\|}(\Gamma)\approx\pm 0.0889$ and $K^{\rm crit}_{\|}(K)\approx\pm 0.0551$. Both of these values are larger than the $K=0.02$, which we chose to show the hybridization of the edge modes. 

For$K_\|=0$, the critical value of the inter-dimer nematic interaction is $K'^{\rm crit}_{\|}=\mp\tilde{J'}\pm\sqrt{3}\tilde{D}\pm\sqrt{\tilde{D}^2+\tilde{J'}{}^2}$. The lower gap closes at the $\Gamma$ point when $\tilde{D}\approx(D'+D'')(\frac{J'}{J}+1)$ and $ \tilde{J'}\approx J'-\frac{J'^2}{2 J}+\frac{(D'+D'')^2}{2 J}$ and the gap between the top bands closes at the $K(K')$ point when $\tilde{D}\approx(D'-2D'')(1-\frac{J'}{2 J})$ and $ \tilde{J'}\approx J'+\frac{J'^2}{4 J}-\frac{(D'-2D'')^2}{4 J}$. For the parameter values of Fig.~\ref{fig:nematic_open} (d)--(e), the $K'^{\rm crit}_{\|}(\Gamma)\approx\pm 0.0197$ and $K'^{\rm crit}_{\|}(K)\approx\pm 0.0312$, both of which are exceeding $K'=0.01$ that we chose.

\bibliographystyle{apsrev4-1}
\bibliography{main}

\end{document}